\documentclass[12pt,aps,pre,preprint,showpacs,showkeys]{revtex4}

\usepackage{graphicx}
\usepackage{subfigure}
\usepackage{amsmath}
\usepackage{tikz}
\usepackage{setspace}
\usepackage{amssymb}
\usepackage{bm}

\begin{document}

\title{New ordered phase in geometrically frustrated generalized $XY$ model}
\author{M. Lach},
\author{M. \v{Z}ukovi\v{c}}
\email{milan.zukovic@upjs.sk}
\affiliation{Institute of Physics, Faculty of Science, P.J. \v{S}af\'arik University, Park Angelinum 9, 041 54 Ko\v{s}ice, Slovakia}
\date{\today}

\begin{abstract}
Critical properties of a geometrically frustrated generalized $XY$ model with antiferromagnetic (AFM) and third-order antinematic (AN3) couplings on a triangular lattice are studied by Monte Carlo simulation. It is found that such a generalization leads to a phase diagram consisting of three different quasi-long-range ordered (QLRO) phases. Compared to the model with the second-order antinematic (AN2) coupling, besides the AFM and AN3 phases which appear in the limits of relatively strong AFM and AN3 interactions, respectively, it includes an additional complex canted antiferromagnetic (CAFM) phase. It emerges at lower temperatures, wedged between the AFM and AN3 phases, as a result of the competition between the AFM and AN3 couplings, which is absent in the model with the AN2 coupling. The AFM-CAFM and AN3-CAFM phase transitions are concluded to belong to the weak Ising and weak three-state Potts universality classes, respectively. Additionally, all three QLRO phases also feature true LRO of the standard and generalized chiralities, which both vanish simultaneously at second-order phase transitions with non-Ising critical exponents and the critical temperatures slightly higher than the magnetic and nematic order-disorder transition temperatures. 
\end{abstract}



\maketitle

\section{Introduction}
Despite the rigorously proven absence of any true long-range ordering~\cite{Mermin_Wagner_1966}, the standard two-dimensional $XY$ model is, nevertheless, known to exhibit the Berezinskii-Kosterlitz-Thouless (BKT) phase transition~\cite{BKT1, BKT2}. This infinite-order phase transition is driven by the unbinding of topological defects in the form of vortices. At low temperature, the integer valued vortices are all joined in vortex-antivortex pairs, resulting in an algebraically decaying spin-spin correlation function and the so-called quasi-long-range order (QLRO). At the BKT critical temperature, these pairs unbind, the correlation function decay becomes exponential and the system becomes completely disordered.

The model with antiferromagnetic (AFM) interactions on a non-bipartite, such as triangular, lattice becomes geometrically frustrated. It has been intensively studied in relation with the possibility of separate phase transitions to the vector chiral LRO and the magnetic QLRO phases (spin-chirality decoupling) and the corresponding universality classes~\cite{Miyashita_1984, Lee_1986, Lee_1998, Korshunov_2002, Hasenbusch_2005, Obuchi_2012}. 

The standard $XY$ model can be generalized by the inclusion of higher-order harmonics, leading to the Hamiltonian
\begin{equation}
\label{Generalized_Hamiltonian}
\mathcal{H} = -J_1\sum_{\langle i,j\rangle} \cos(\phi_i - \phi_j) - J_q\sum_{\langle i,j\rangle} \cos[ q(\phi_i - \phi_j)],
\end{equation}   
where $\phi_i  \in [0,2\pi]$ represents the $i$-th site spin angle in the $XY$ plane, $J_1$ and $J_q$ are exchange interaction parameters and $\langle i,j\rangle$ denotes the sum over nearest-neighbor spins. The first term $J_1$ is a usual magnetic, i. e. ferromagnetic, FM, ($J_1 > 0$)  or AFM ($J_1 < 0$) coupling, while the second term $J_q$ represents a generalized nematic, Nq, ($J_q > 0$) or antinematic, ANq, ($J_q < 0$) interaction.  

The model (\ref{Generalized_Hamiltonian}) with $q = 2$ has been studied for the non-frustrated FM-N2 interactions (both $J_1$ and $J_2$ positive)~\cite{Lee_Grinstein_1985, Korshunov_1985, Sluckin_1988, Carpenter_1989, hlub08, QI_2013, Hubscher_2013} and more recently also for the frustrated AFM - AN2 interactions (both $J_1$ and $J_2$ negative)~\cite{Park-nematic}. In both cases, this generalization led, for sufficiently large ratio $J_2/J_1$, to a new phase transition between the magnetically and nematically ordered phases belonging to the Ising universality class and in the frustrated case, additionally to a separate chiral phase transition above the BKT transition line~\cite{Park-nematic}. On the other hand, in the model on a bipartite square lattice with a frustration parameter, it has been found that for the magnetic and nematic couplings of comparable strengths, the chirality becomes disordered before the BKT transition line~\cite{qin09}.

Even more interesting is the case when $J_1$ and $J_2$ compete. The ground-state phase diagrams of Heisenberg and XY models with different types of bilinear and biquadratic exchange interactions with square- and rhombic symmetries produced a variety of different phases~\cite{hayd10}. Theoretical investigations of the model on a square lattice with the geometrically nonfrustrated but mutually competing FM - AN2 interactions revealed the existence of a new phase at very low temperatures~\cite{dian11,zuko19}. Geometrically frustrated models with the magnetic and nematic couplings having opposite signs on a triangular lattice have also found their interdisciplinary applications for modeling of DNA packing~\cite{gras08} and structural phases of cyanide polymers~\cite{zukovic_frustrated_2016}.

Furthermore, a recent series of papers \cite{Poderoso-2011, Canova-2014, Canova-2016} has shown that increasing the order of the couplings to $q > 2$ on a square lattice with FM - Nq interactions, can lead to drastic changes of the phase diagram topology, featuring new phases and phase transitions belonging to a variety of universality classes. This pointed to a rather surprising lack of universality in systems showing the same $\phi \rightarrow \phi + 2\pi$ symmetry. 

Motivated by the above theoretical considerations and by the recent investigations of the ground-state properties of such a model with geometrical frustration~\cite{zukovic_frustrated_2016}, which suggested an interesting physical behavior with potential interdisciplinary applications, in the present study we focus on the critical behavior of the geometrically frustrated model on a triangular lattice with the antiferromagnetic and generalized third-order antinematic interactions. In such a model, besides the phenomenon of geometrical frustration the two interactions compete, which leads to a novel critical behavior. 

\section{Model and Methods} 
We consider the model (\ref{Generalized_Hamiltonian})  for $q = 3$ on the triangular lattice and the interaction parameters $J_1<0$ and $J_3<0$  in the form $J_1 = -\Delta$, $J_3 = \Delta-1$, with $\Delta  \in [0,1]$ to cover the interactions between the pure AN3 $(\Delta = 0)$ and the pure AFM $(\Delta = 1)$ limits. 

Monte Carlo (MC) simulations, based on the standard Metropolis algorithm, implemented on graphical processing units, were employed to simulate the studied system. We considered systems of linear sizes starting from $L = 36$ up to 384, with periodic boundary conditions. Occasional checks were made on larger systems with $L = 768$. The simulations were carried out in two different modes. The first mode was used to probe the whole relevant temperature range from $T = 0.01$, which approximates ground-state conditions, up to $T = 0.52$, corresponding to the paramagnetic phase. At each temperature step $6\times10^5 - 2.5\times10^6$ MC sweeps (more sweeps for large lattice sizes) were used with typically about $20\%$ discarded for equilibration. The second mode was used to determine the critical behavior with up to $1.6\times10^7$ MC sweeps per temperature step and configurational averaging of up to $100$ independent runs. The rather large number of MC sweeps was necessary due to long autocorrelation times, particularly at low and transition temperatures.

The following quantities were calculated: the internal energy per spin
\begin{equation}
\label{eq:ene}
 e =\frac{\langle \mathcal{H} \rangle}{L^2},
 \end{equation}
 the specific heat per spin
\begin{equation}
\label{eq:spec_heat}
c = \frac{\langle \mathcal{H}^2\rangle  -  \langle \mathcal{H}\rangle^2}{T^2 L^2},
\end{equation}
 the magnetic $(m_1)$ and generalized nematic $(m_3)$ order parameters
\begin{equation}
\label{eq:magnetiz}
m_k = \frac{\langle M_k\rangle}{L^2} = \frac{1}{L^2} \left\langle  \sqrt{3 \sum_{\alpha = 1}^3 \textbf{M}^2_{k\alpha}} \right\rangle, k = 1, 3; \alpha = 1, 2, 3;
\end{equation}
where $\textbf{M}_{k\alpha}$ is the $\alpha$-th sublattice order parameter vector given by
\begin{equation}
\label{eq:mag_mom}
\textbf{M}_{k\alpha} =  \left(\sum_{i \in \alpha} \cos(k\phi_{\alpha i}),\sum_{i \in \alpha} \sin(k\phi_{\alpha i})\right),
\end{equation}   
and finally, the standard ($\kappa_1$) and generalized ($\kappa_3$) staggered chiralities
\begin{equation}
\label{eq:chiral}
\kappa_k = \frac{\langle K_k\rangle}{L^2}= \frac{1}{2L^2}\left\langle \left| \sum_{p^+\in\bigtriangleup} \kappa_{kp^+} -  \sum_{p^-\in\bigtriangledown} \kappa_{kp^-} \right| \right\rangle, k = 1, 3;
\end{equation}
where $\kappa_{kp^+}$ and $\kappa_{kp^-}$ are the local generalized chiralities for each elementary plaquette of upward and downward triangles, respectively, defined by:
\begin{equation}
\label{eq:chiral_lok}
\kappa_{kp} = 2\{\sin[k(\phi_2-\phi_1)] + \sin[k(\phi_3-\phi_2)]+ \sin[k(\phi_1-\phi_3)]\}/3\sqrt{3}.
\end{equation}
The susceptibilities of the order parameters can be defined in the following way:
\begin{equation}
\label{eq:susc}
\chi_o = \frac{1}{TL^2} (\langle \mathcal{O}^2\rangle  -  \langle \mathcal{O}\rangle^2), \mathcal{O} = M_1, M_3, K_1, K_3.
\end{equation}
It is also useful to calculate the following quantities:
\begin{equation}
\label{eq:log_der}
D_{lk} = \frac{\partial}{\partial \beta} \ln\langle \mathcal{O}^l_k\rangle = \frac{\langle \mathcal{O}^l_k  \mathcal{H}\rangle}{\langle \mathcal{O}^l_k \rangle} - \langle \mathcal{H}\rangle, \mathcal{O} = M_k, K_k; l = 1, 2; k = 1, 3.
\end{equation}
At standard second-order phase transitions, the order parameters~(\ref{eq:magnetiz}) and~(\ref{eq:chiral}) and the extreme values of the quantities~(\ref{eq:susc}) and~(\ref{eq:log_der}) scale with the system size as
\begin{equation}
\label{eq:fss_o}
o(L) \propto L^{-\beta / \nu},
\end{equation}
\begin{equation}
\label{eq:fss_chi}
\chi_{o,max}(L) \propto L^{\gamma / \nu},
\end{equation}
\begin{equation}
\label{eq:fss_D}
D_{l,k,max}(L) \propto L^{1 / \nu}.
\end{equation}
Within the QLRO phases the respective order parameters scale with the system size as
\begin{equation} 
\label{eq:fss_o_bkt}
o(L) \propto L^{-\eta_o (T)}, 
\end{equation}
where $\eta_o (T)$ is the temperature dependent critical exponent of the correlation function for the order parameters $o = m_1$ and $m_3$. 

\section{Results}
\subsection{Order Parameters and Phase Diagram}
\begin{figure}
\centering
\subfigure{\includegraphics[scale=0.4,clip]{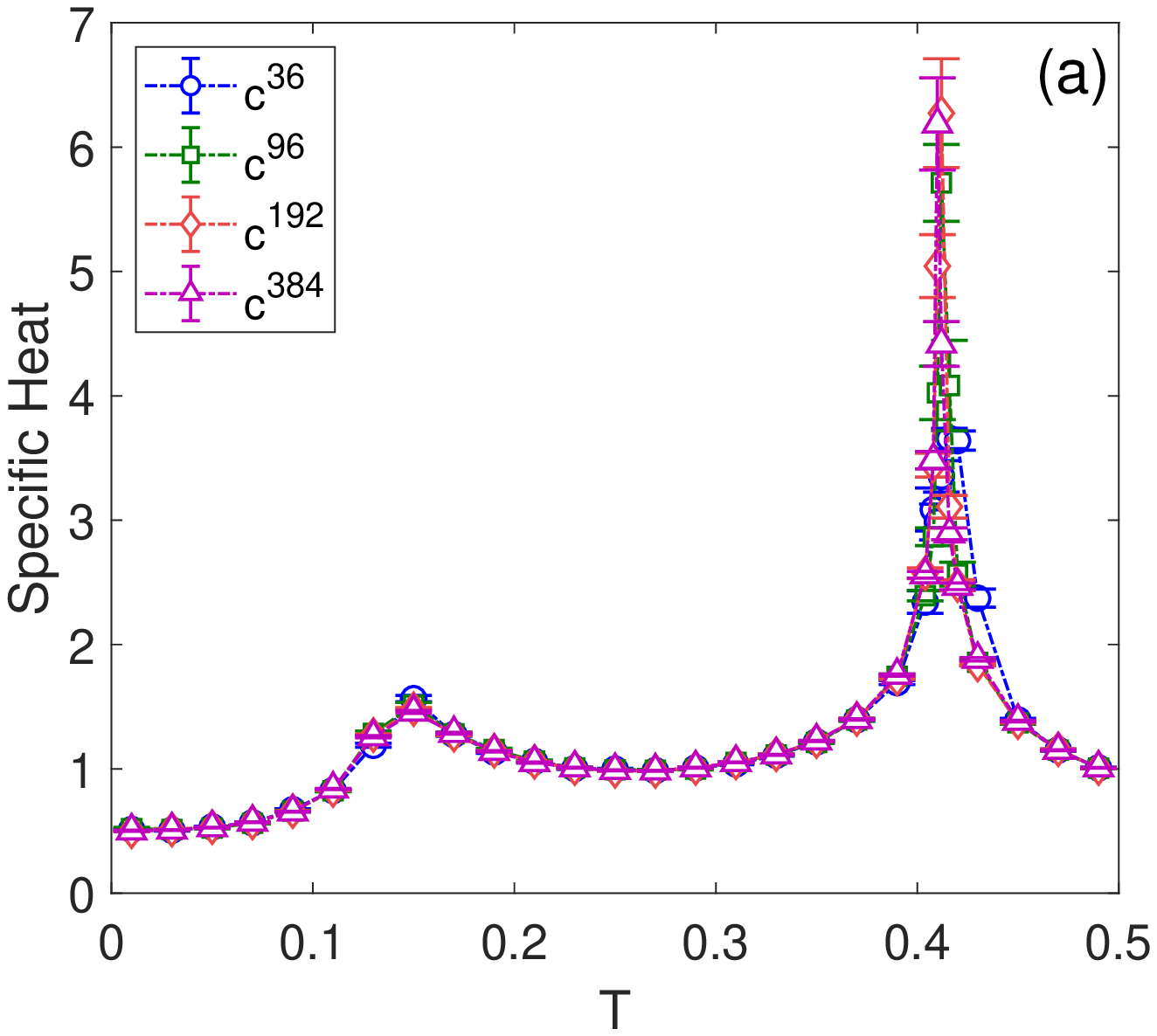}\label{fig:Cx02}}
\subfigure{\includegraphics[scale=0.4,clip]{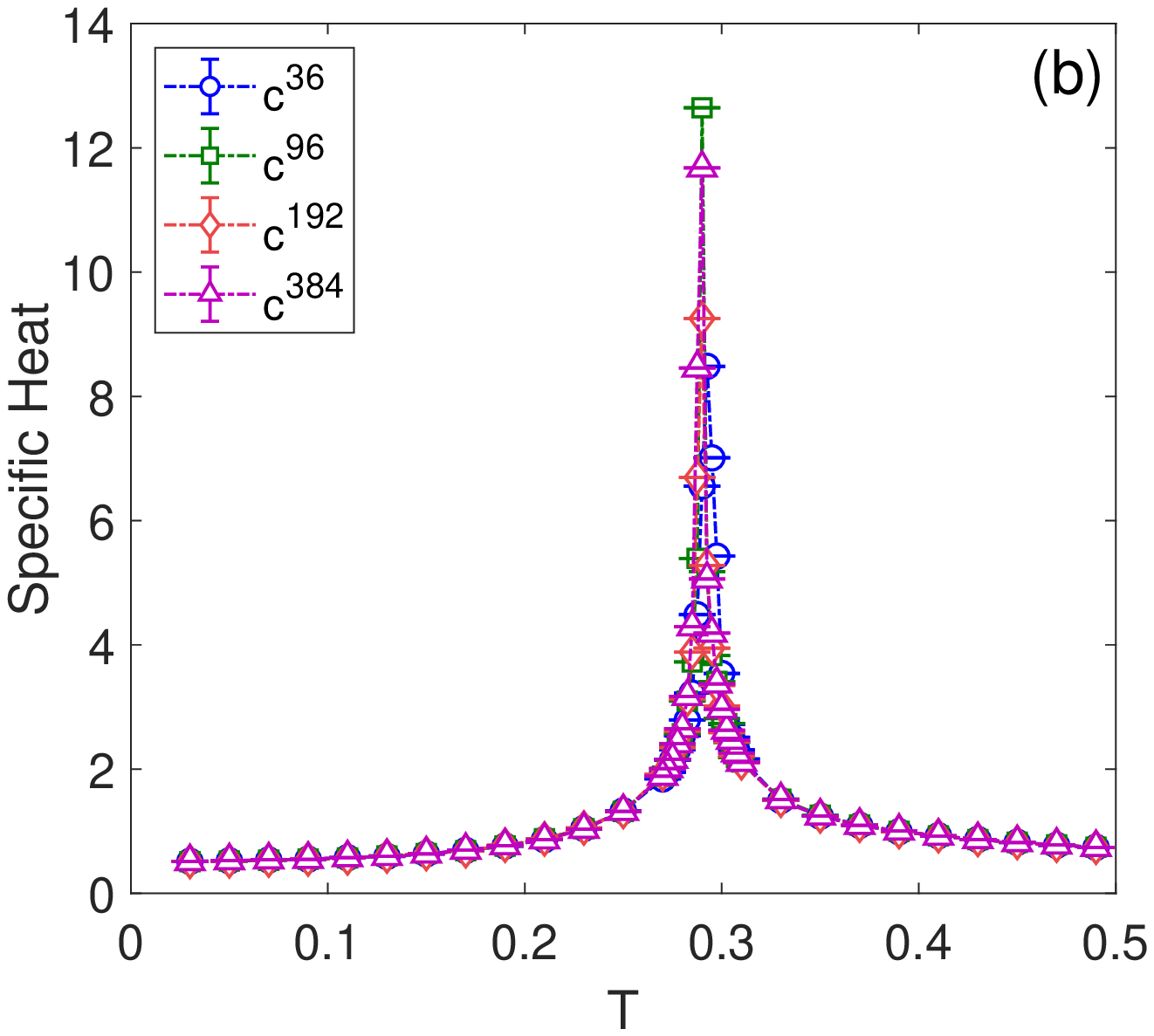}\label{fig:Cx0555}}
\subfigure{\includegraphics[scale=0.4,clip]{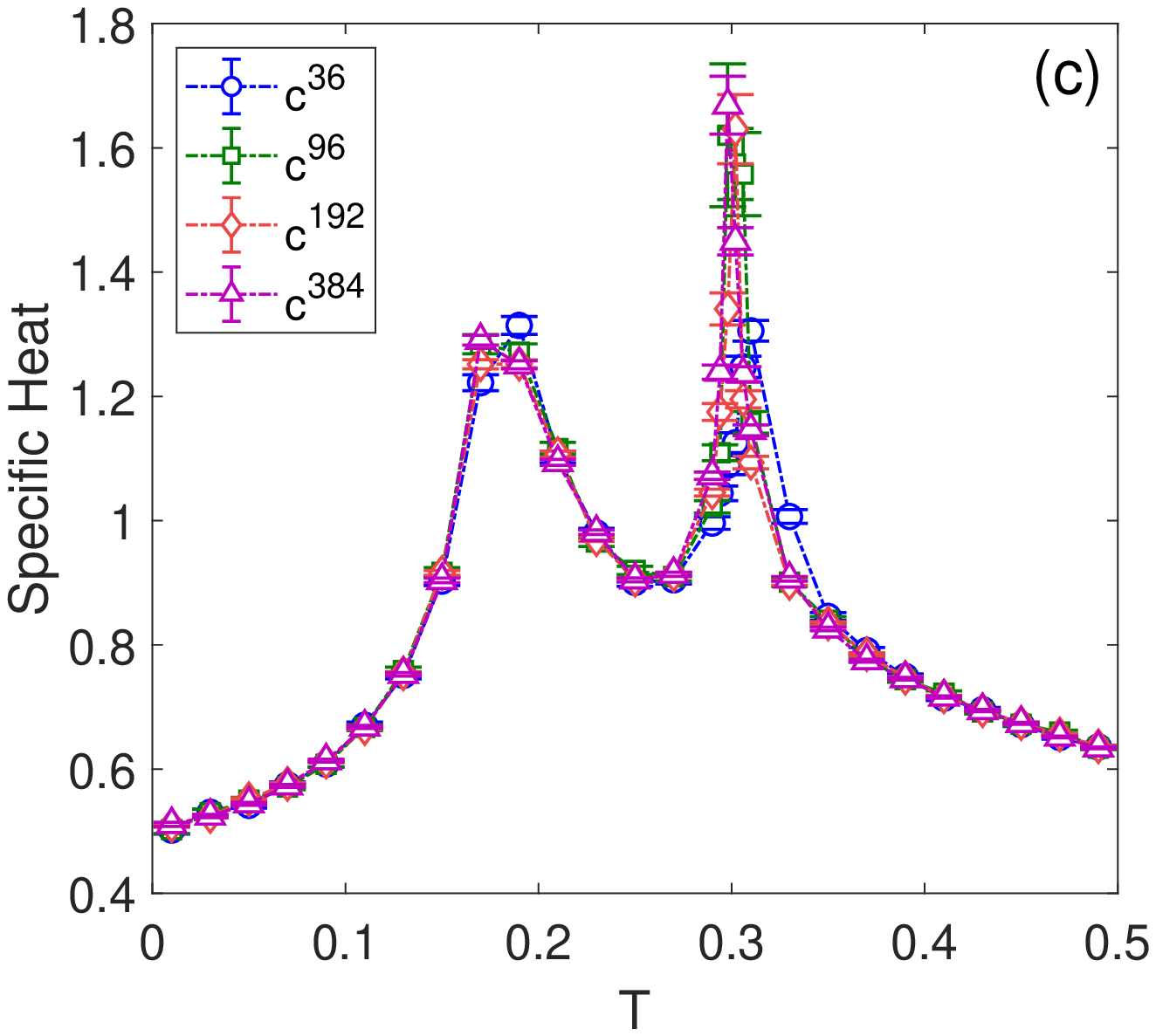}\label{fig:Cx08}}\\
\subfigure{\includegraphics[scale=0.4,clip]{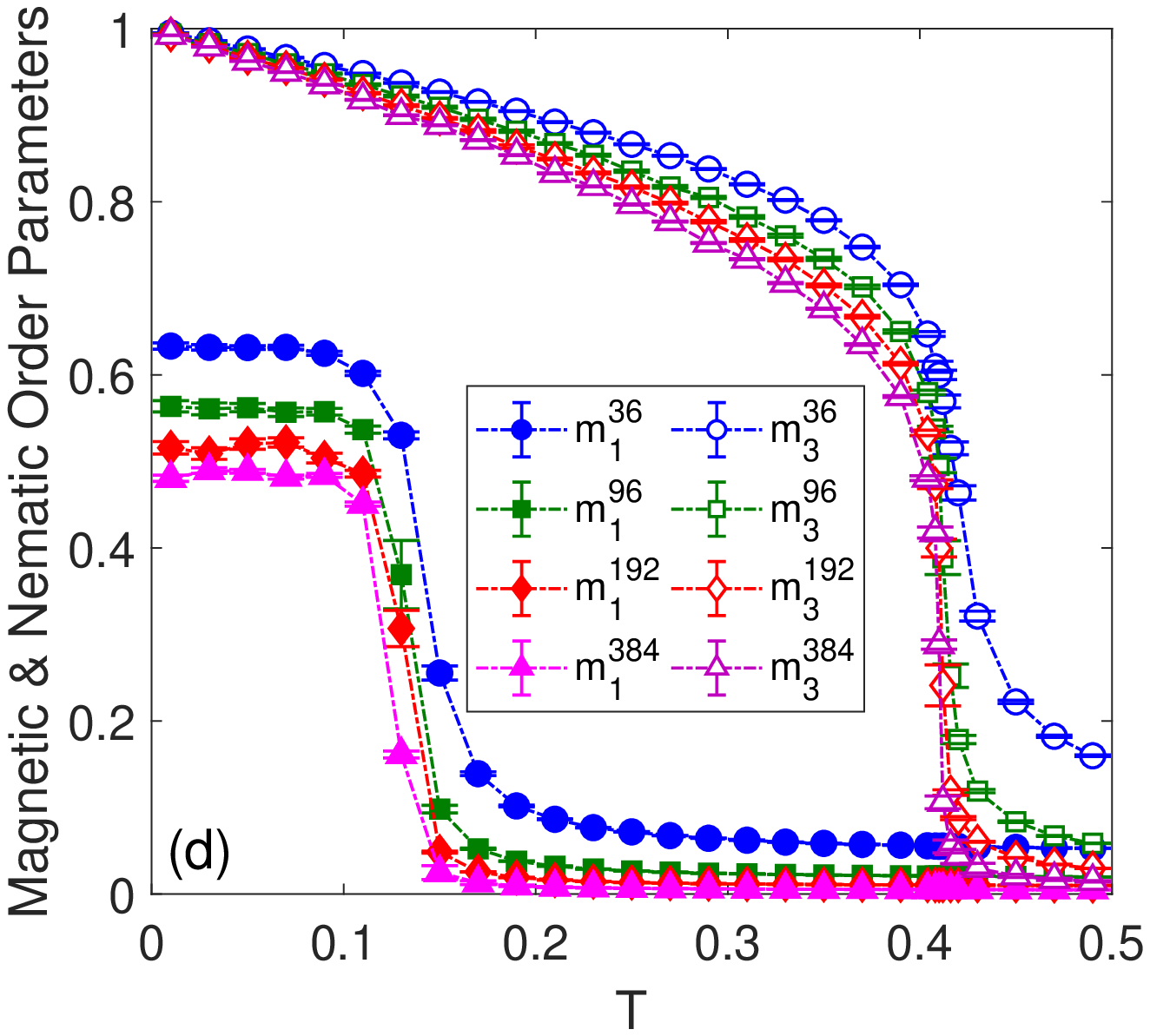}\label{fig:m13x02}}
\subfigure{\includegraphics[scale=0.4,clip]{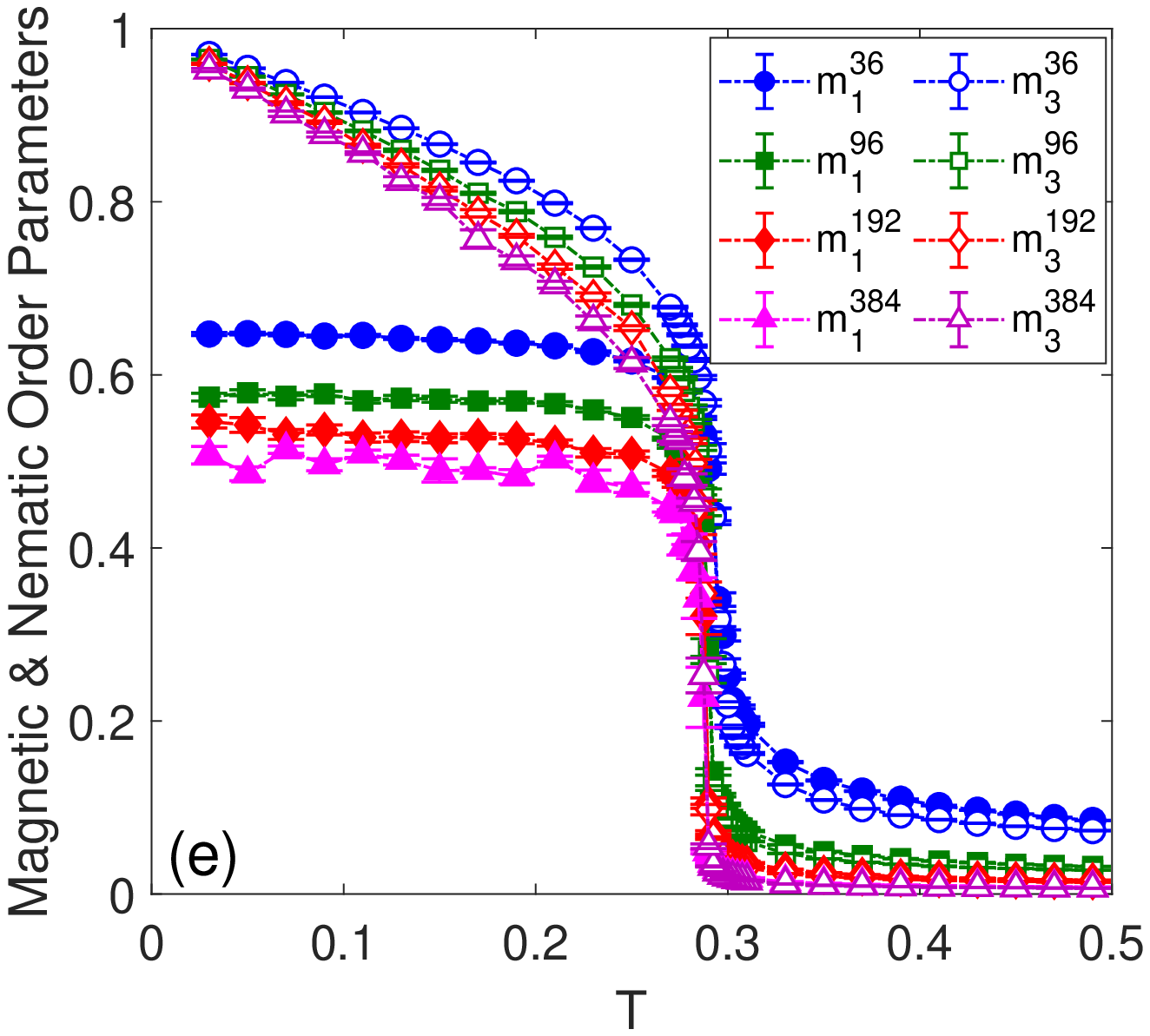}\label{fig:m13x0555}}
\subfigure{\includegraphics[scale=0.4,clip]{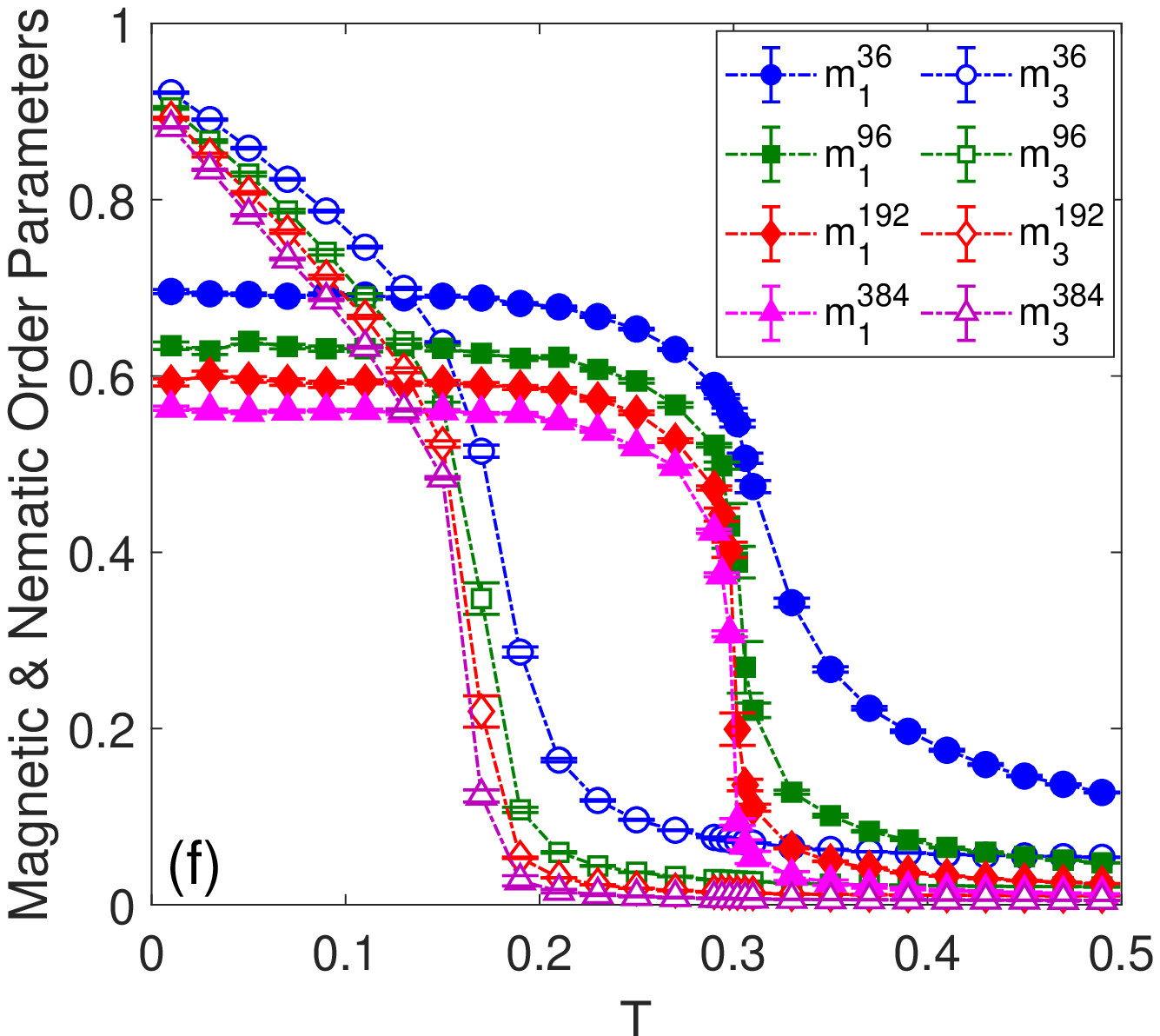}\label{fig:m13x08}}\\
\subfigure{\includegraphics[scale=0.4,clip]{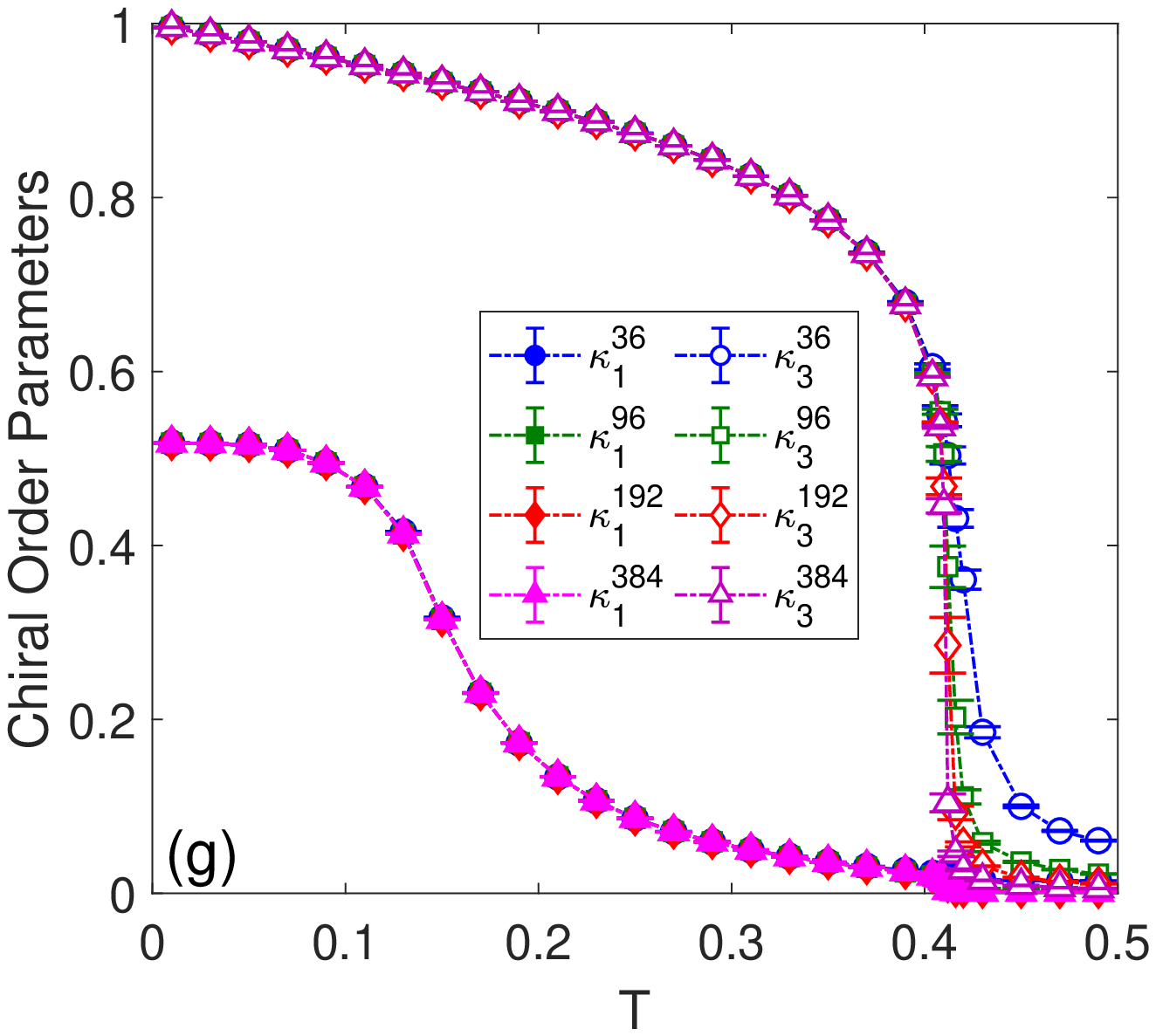}\label{fig:kappa13x02}}
\subfigure{\includegraphics[scale=0.4,clip]{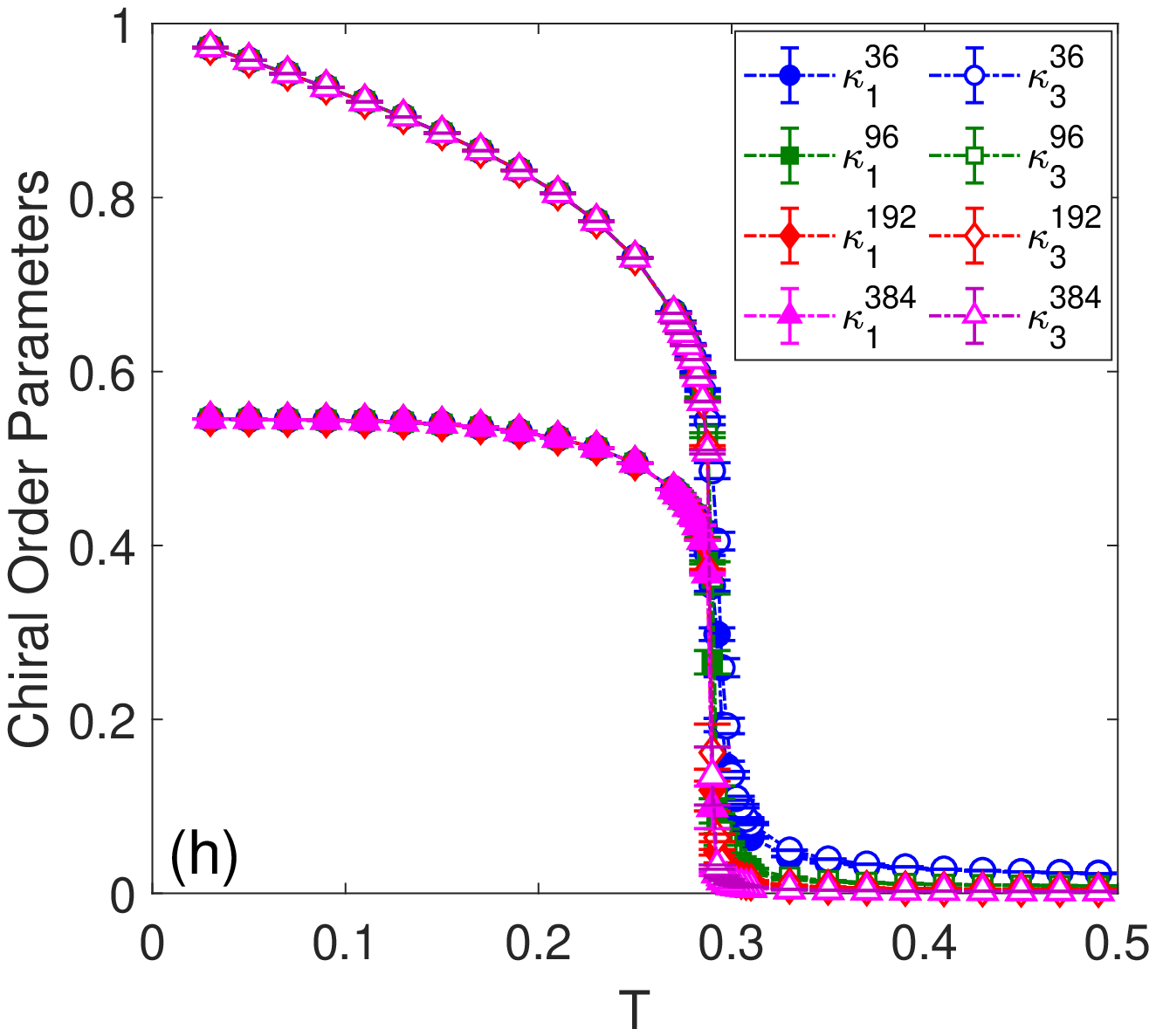}\label{fig:kappa13x0555}}
\subfigure{\includegraphics[scale=0.4,clip]{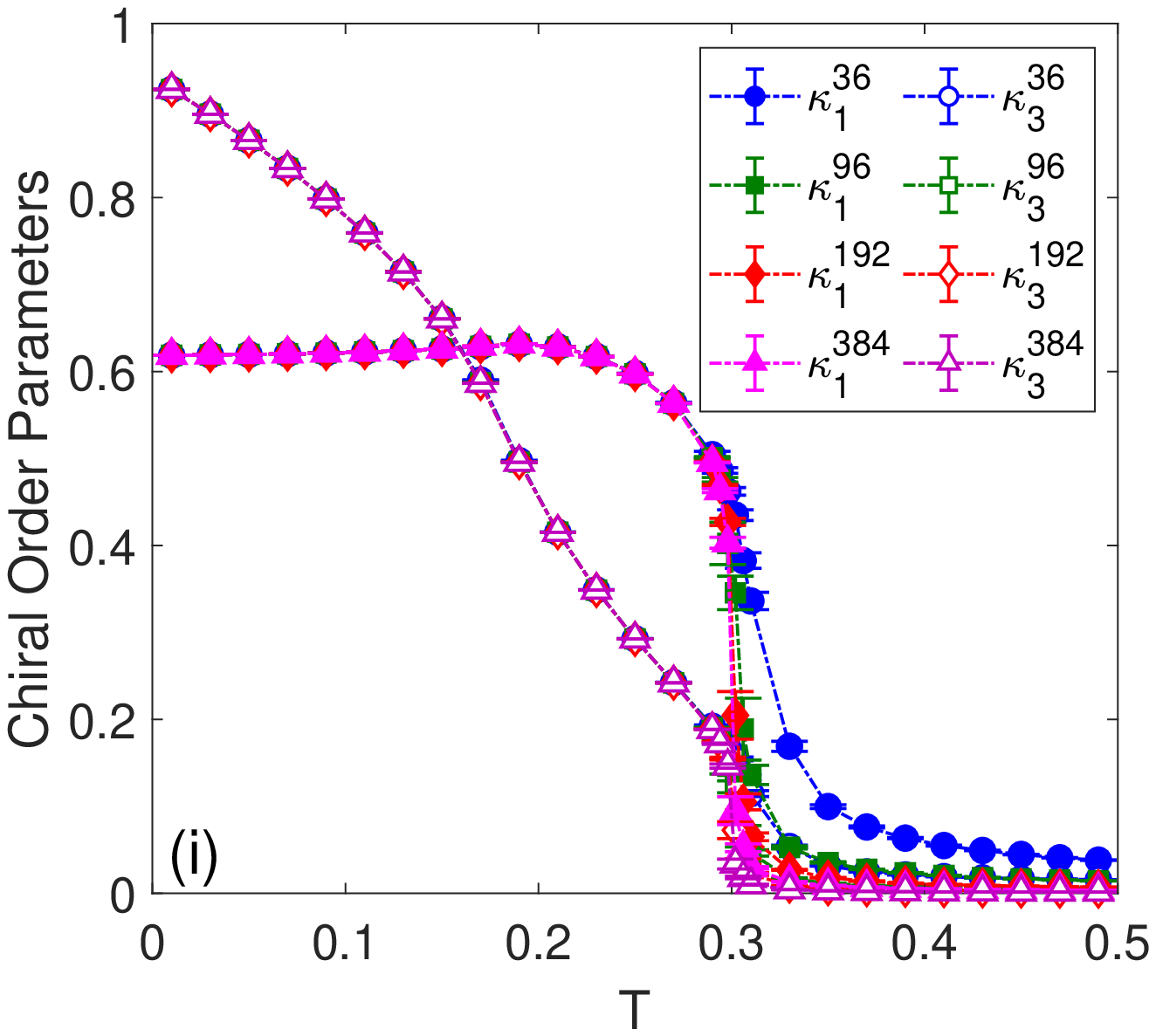}\label{fig:kappa13x08}}
\caption{Temperature dependencies of the specific heat $c$ (upper row), magnetic and generalized nematic order parameters $m_1$ and $m_3$ (middle row), and standard and generalized chiral order parameters $\kappa_1$ and $\kappa_3$ (lower row), for three representative values of $\Delta=0.2$ (first column), $\Delta=0.555$ (second column), and $\Delta=0.8$ (third column), and different sizes $L$ from 36 to 384.}
\label{fig:order}
\end{figure}
The ground-state investigations of the present model~\cite{zukovic_frustrated_2016} led to the conclusion that by inclusion of even a relatively small value of the $J_3$ interaction, the chiral AFM state of the pure $J_1<0$ and $J_3=0$ model, characterized by the phase angles $\Delta \phi =\pm 2\pi/3$, qualitatively changes. In particular, within $0 < \Delta  \lesssim 0.997$ it shows a peculiar canted AFM (CAFM) phase in which pairs of neighboring spins on each triangular plaquette form angles with $\Delta$-dependent values in such a way that two neighbors are oriented almost parallel with respect to each other and almost antiparallel with respect to the third one. Thus, at finite temperatures one would expect that, at least in the vicinity of the limiting values of $\Delta$, there might be two phase transitions: first from the paramagnetic to the AN3 (AFM) phase for $\Delta \gtrsim 0$ ($\Delta \lesssim 0.997$), followed by the second one to CAFM at lower temperature.

Temperature dependencies of the measured quantities, plotted in Fig.~\ref{fig:order} for representative values of $\Delta$ and various lattice sizes, indeed indicate such a behavior. In particular, the specific heat measurements show two peaks for $0.0 < \Delta <0.5$ and $0.6 < \Delta \lesssim 0.997$  (Figs.~\ref{fig:Cx02} and~\ref{fig:Cx08}), indicating two phase transitions. On the other hand, in the case of roughly equal interactions ($0.5\leq \Delta\leq 0.6$), there is only a single peak (Fig.~\ref{fig:Cx0555}), suggesting the presence of only one phase transition. 

A better picture of the nature of the respective phases can be obtained from the temperature variation of the magnetic $(m_1)$ and generalized nematic $(m_3)$ order parameters, plotted in the middle row of Fig.~\ref{fig:order}. They show that for $0 < \Delta <0.5$ the magnetic order vanishes at temperatures lower than the nematic one (Fig.~\ref{fig:m13x02}). The gap between the two transition temperatures shrinks with the increasing $\Delta$ and in the vicinity of $\Delta =0.555$, it disappears completely (Fig.~\ref{fig:m13x0555}). Thus at $\Delta \approx 0.555$ the transition from the paramagnetic phase is to neither AN3 nor AFM phases but straight to the CAFM phase. Furthermore, the mutual competition of the AFM and AN3 couplings pushes the transition temperature to lower values, for intermediate $\Delta$ corresponding to only $50-60\%$ of the value of the pure $XY$ model ($\Delta=1$). For $0.6 < \Delta \lesssim 0.997$ the order of the respective transitions is reversed, i.e., the nematic phase vanishes at lower temperatures than the magnetic one (Fig.~\ref{fig:m13x08}). This means that for most values of $\Delta$, there are two distinct QLRO phases: the low-temperature CAFM phase characterized by simultaneous magnetic and nematic ordering and the high-temperature one, with purely generalized antinematic (AN3) or purely antiferromagnetic (AFM) ordering. It is worth mentioning that within the CAFM phase, owing to the geometrical frustration induced by the triangular lattice geometry and the competition between the AFM and AN3 interactions, on approach to zero temperature both order parameters $m_1$ and $m_3$ fail to reach the saturation value, albeit the latter one is very close to it for sufficiently small values of $\Delta$ (see also Ref.~\cite{zukovic_frustrated_2016}). 

\begin{figure}[t!]
\centering
\includegraphics[scale=0.7,clip]{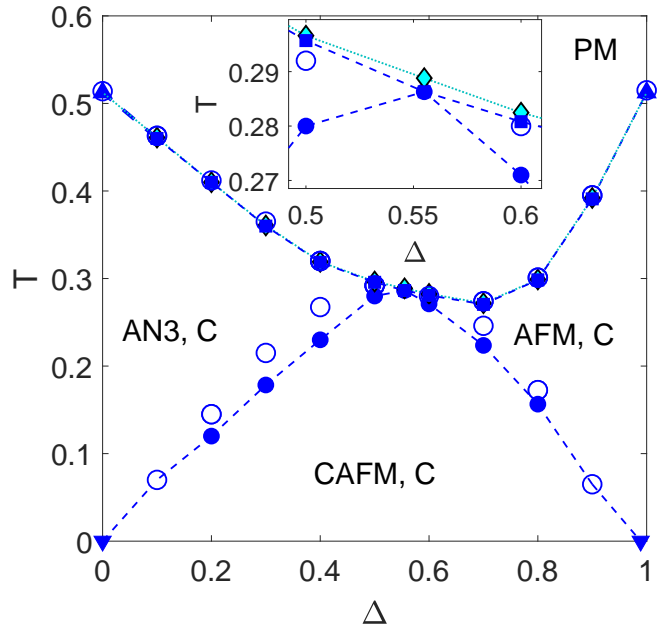}
\caption{Phase diagram in the $\Delta-T$ parameter plane with the generalized antinematic (AN3), canted antiferromagnetic (CAFM), antiferromagnetic (AFM), chiral (C), and paramagnetic (PM) phases. The empty circles represent temperatures corresponding to the maxima of the specific heat for $L=96$, the filled circles result from the data collapse of magnetic or nematic order parameters and susceptibilities, and the filled squares are obtained from the  FSS analysis of the correlation function critical exponent. The cyan diamonds correspond to the chiral transition temperatures obtained from the data collapse of the chiral order parameters and susceptibilities. The down-triangle symbols at the edges represent the limits of the CAFM phase in the ground state (from Ref.~\cite{zukovic_frustrated_2016}) and up-triangles represent the known value of the BKT transition temperature for the standard $XY$ model.}
\label{fig:PD_q3}
\end{figure}

It should be kept in mind that, besides the magnetic and nematic orderings, in the present frustrated model there are also chiral orderings in the system. The last row of Fig.~\ref{fig:order} presents temperature dependencies of the standard ($\kappa_1$) and generalized ($\kappa_3$) staggered chiralities. With the increasing temperature there is an anomalous decrease of the former in the vicinity of the AN3-CAFM phase transition for $0.0 < \Delta <0.5$ and the latter in the vicinity of the AFM-CAFM phase transition for $0.6 < \Delta \lesssim 0.997$. Nevertheless, both remain nonzero up to the temperatures close to the transition to the paramagnetic state where they simultaneously vanish~\footnote{For $\Delta \leq 0.1$ the chirality $\kappa_1$ above the AN3-CAFM transition temperature dropped to very low values, practically indistinguishable from zero, and thus the existence of the $\kappa_1$ chiral LRO in this region is questionable.}. The question whether the transition temperatures of the chiral (C) phase, characterized by a finite values of the chiral order parameters, coincide with those at which the magnetic and nematic phases vanish will be addressed bellow. The resulting phase diagram is presented in Fig.~\ref{fig:PD_q3}.

\subsection{Finite Size Scaling Analysis}
\subsubsection{Order-disorder transitions}
\begin{figure}[t!]
\centering
\subfigure{\includegraphics[scale=0.4,clip]{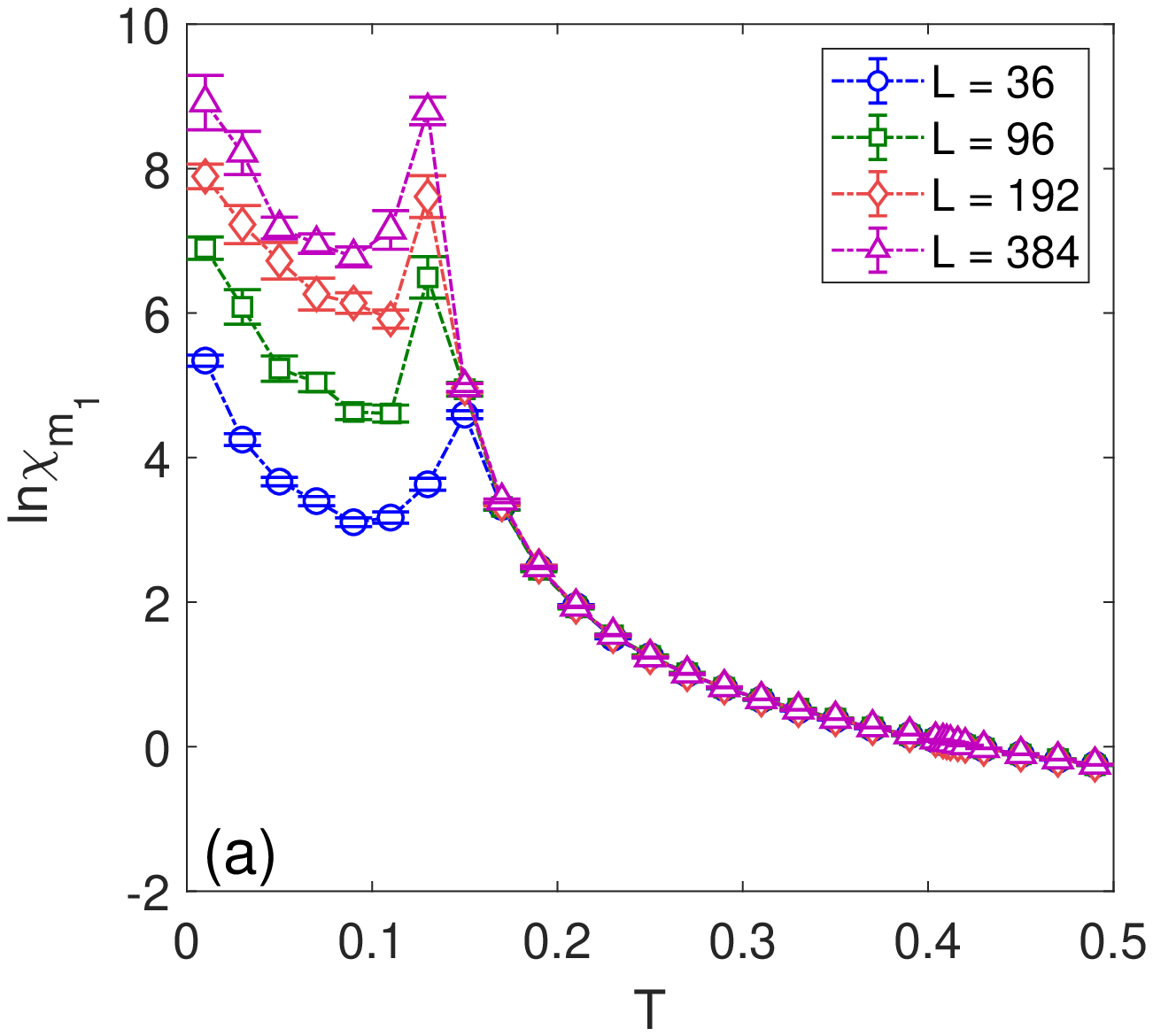}\label{fig:susc_m1x02}}
\subfigure{\includegraphics[scale=0.4,clip]{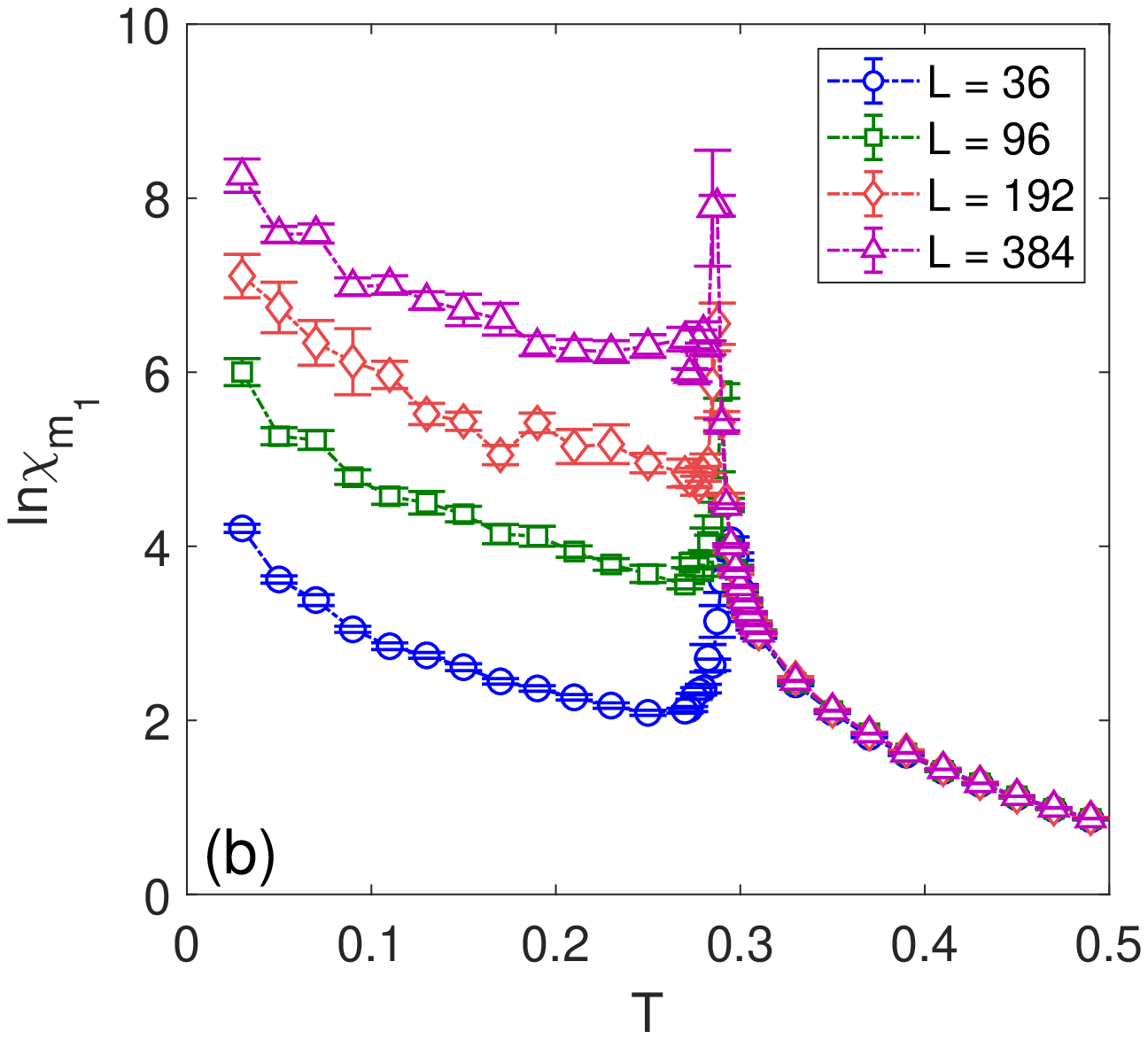}\label{fig:susc_m1x0555}}
\subfigure{\includegraphics[scale=0.4,clip]{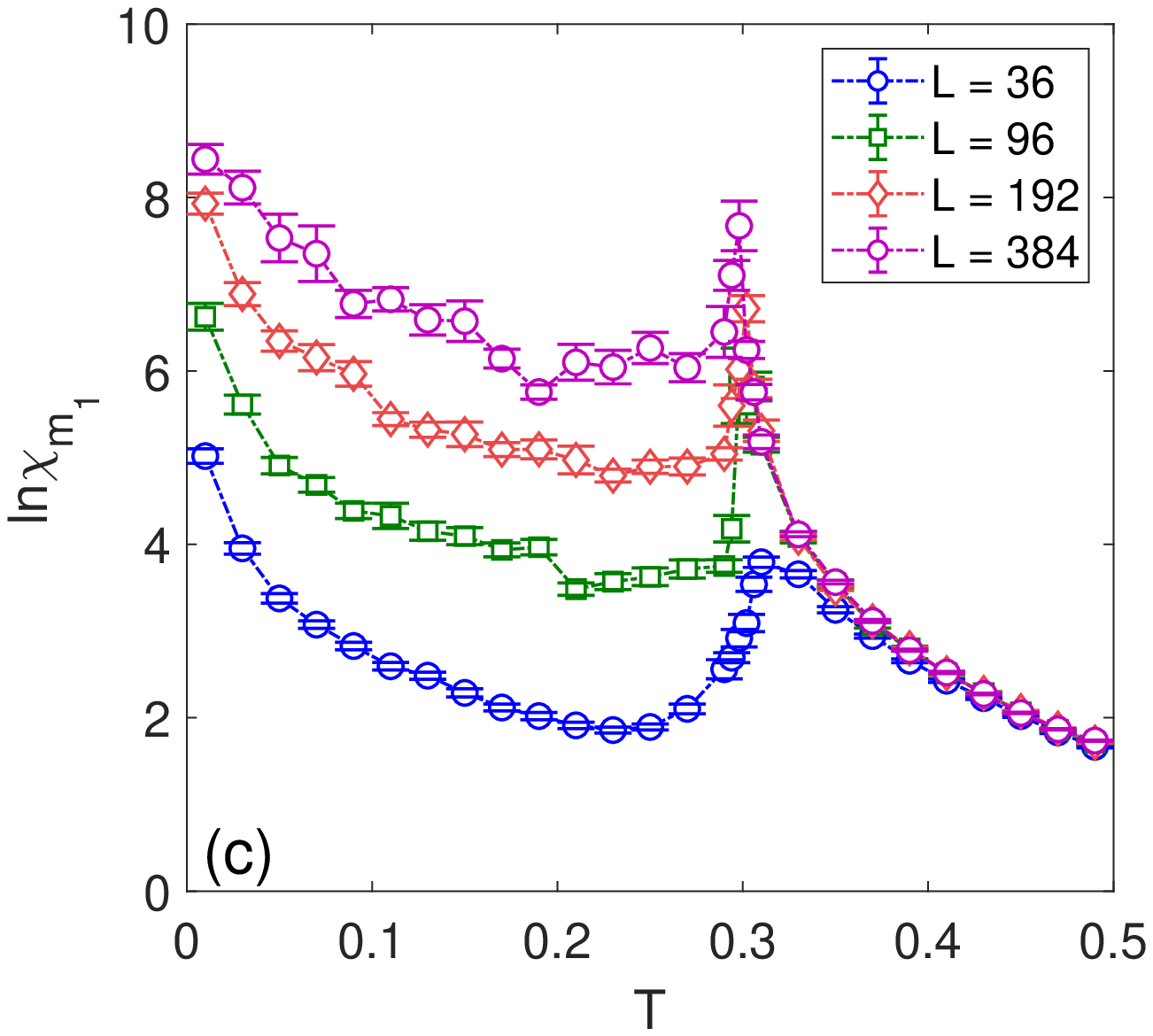}\label{fig:susc_m1x08}}\\
\subfigure{\includegraphics[scale=0.4,clip]{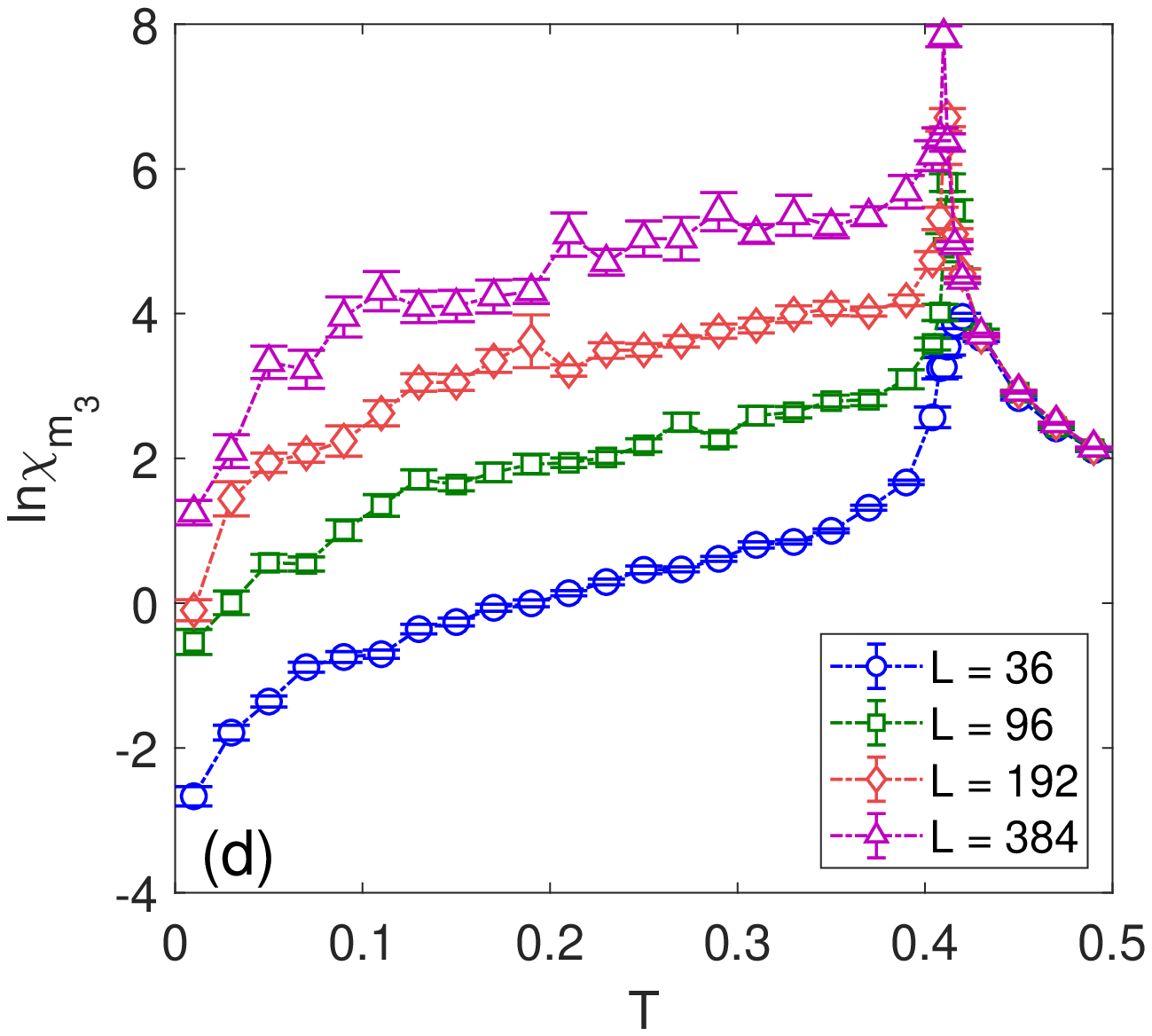}\label{fig:susc_m3x02}}
\subfigure{\includegraphics[scale=0.4,clip]{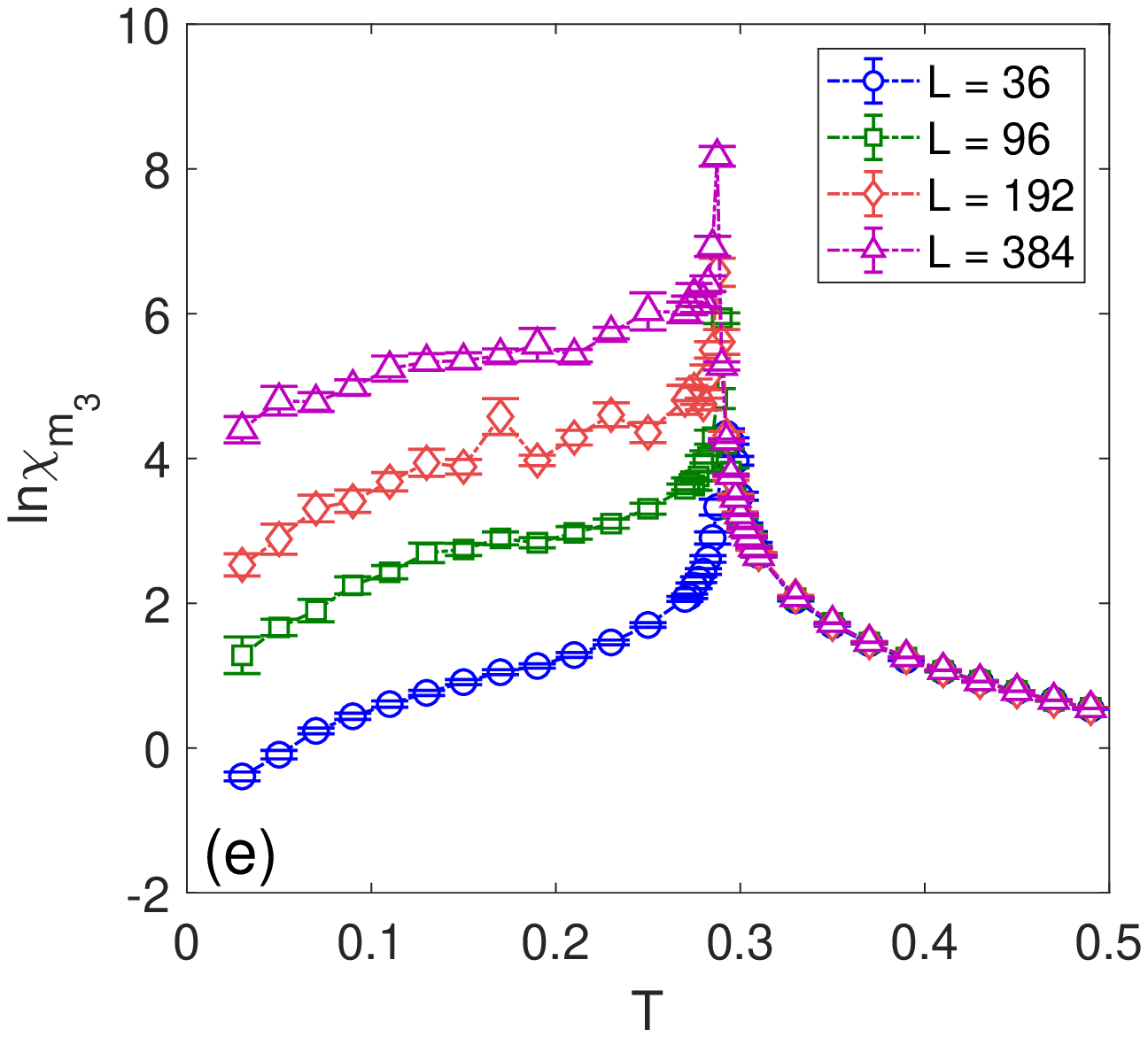}\label{fig:susc_m3x0555}}
\subfigure{\includegraphics[scale=0.4,clip]{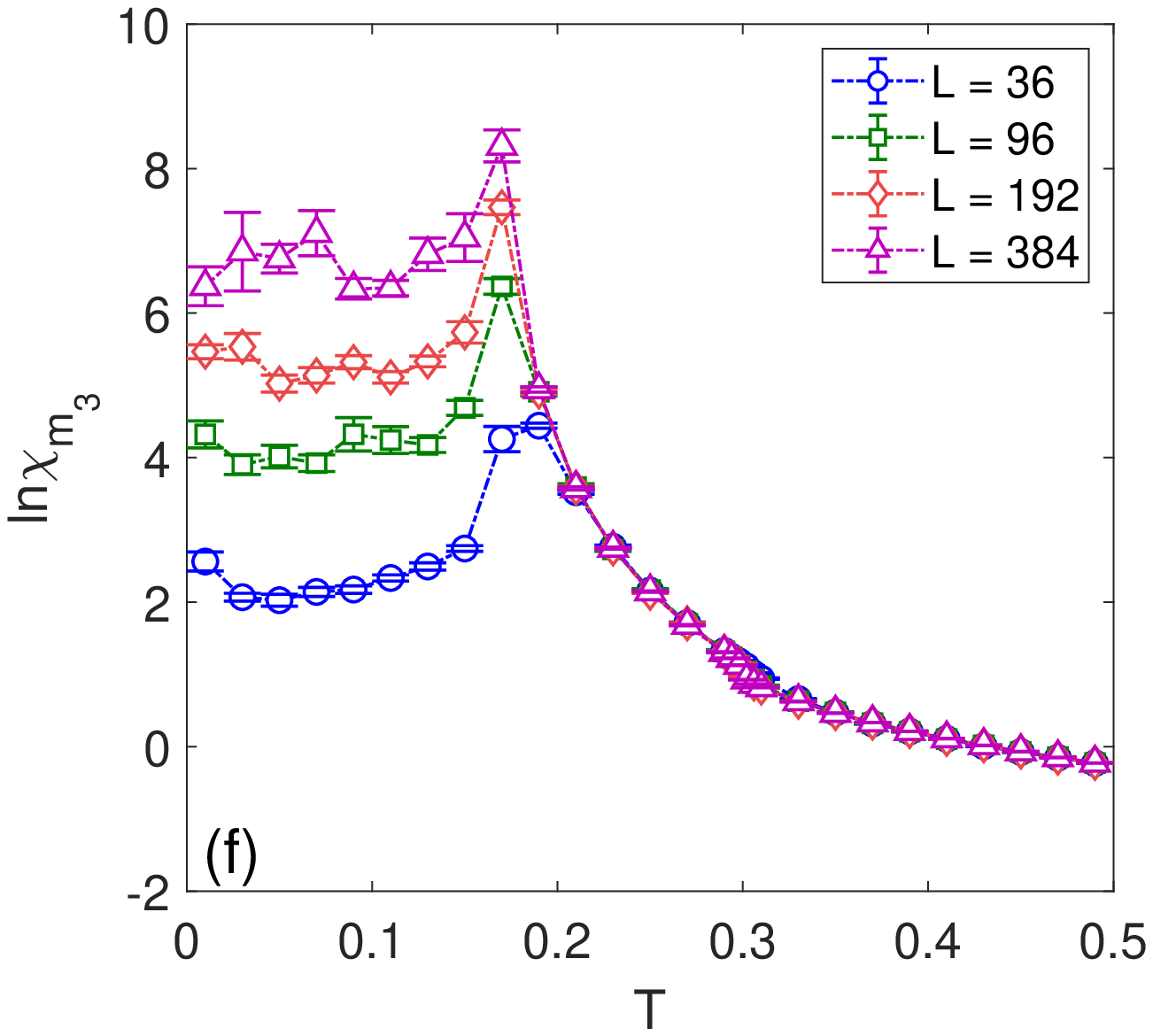}\label{fig:susc_m3x08}}
\caption{Temperature dependencies of the magnetic (upper row) and generalized nematic (lower row) susceptibilities for three representative values of $\Delta=0.2$ (first column), $\Delta=0.555$ (second column), and $\Delta=0.8$ (third column), and different sizes $L$.}
\label{fig:magnetic_susc}
\end{figure}

\begin{figure}[t!]
\centering
\subfigure{\includegraphics[scale=0.4,clip]{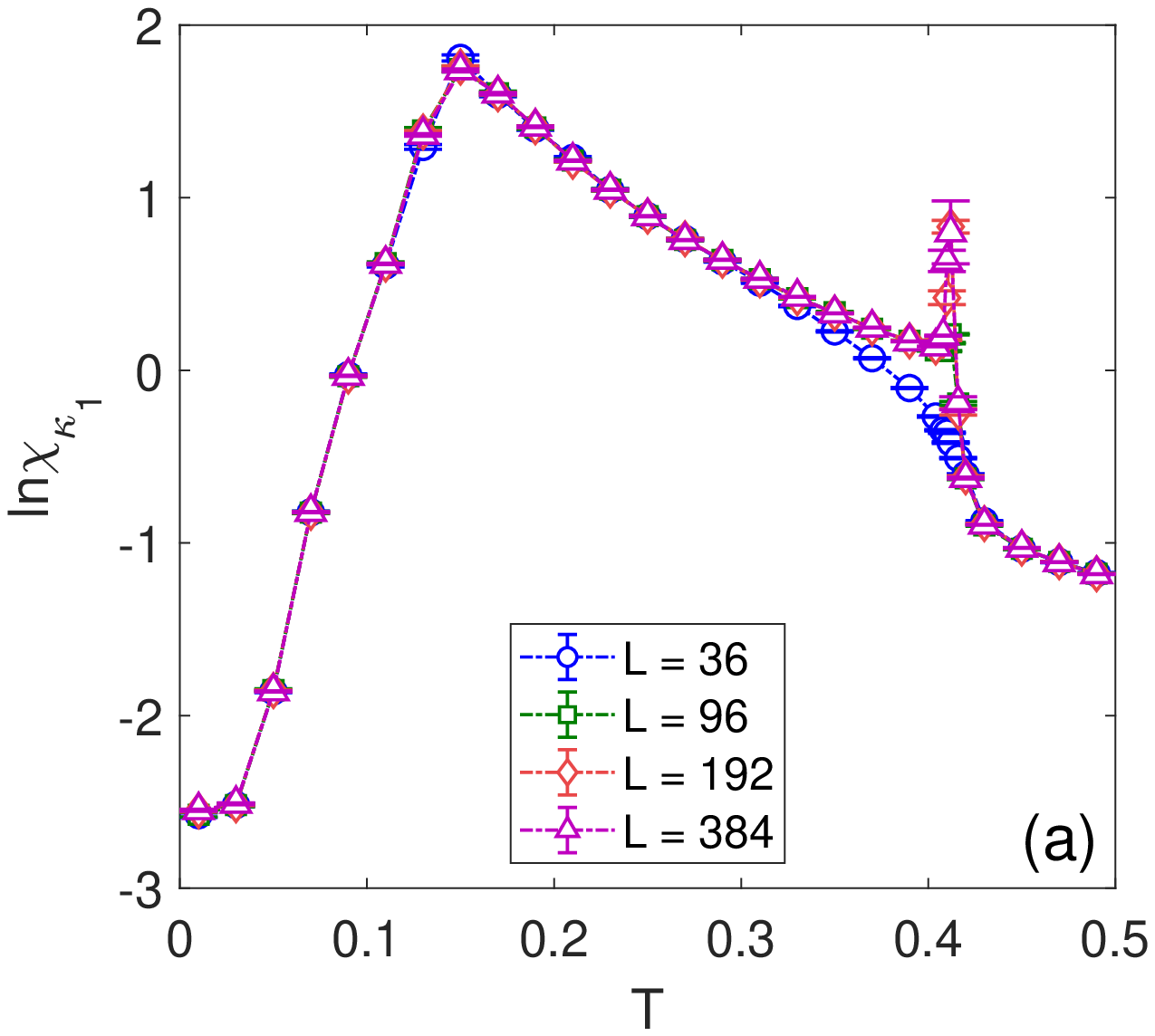}\label{fig:susc_chi1x02}}
\subfigure{\includegraphics[scale=0.4,clip]{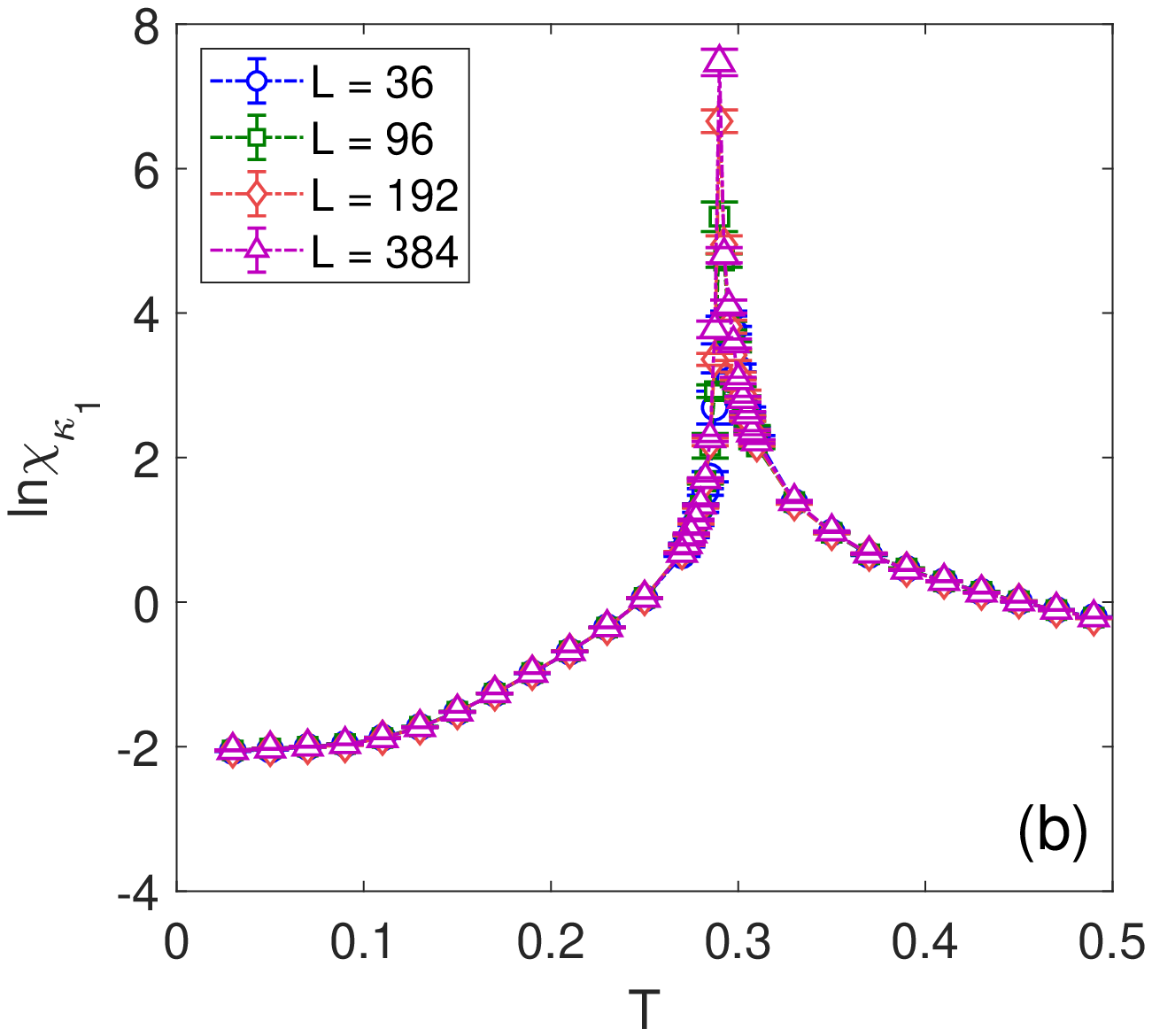}\label{fig:susc_kp1x0555}}
\subfigure{\includegraphics[scale=0.4,clip]{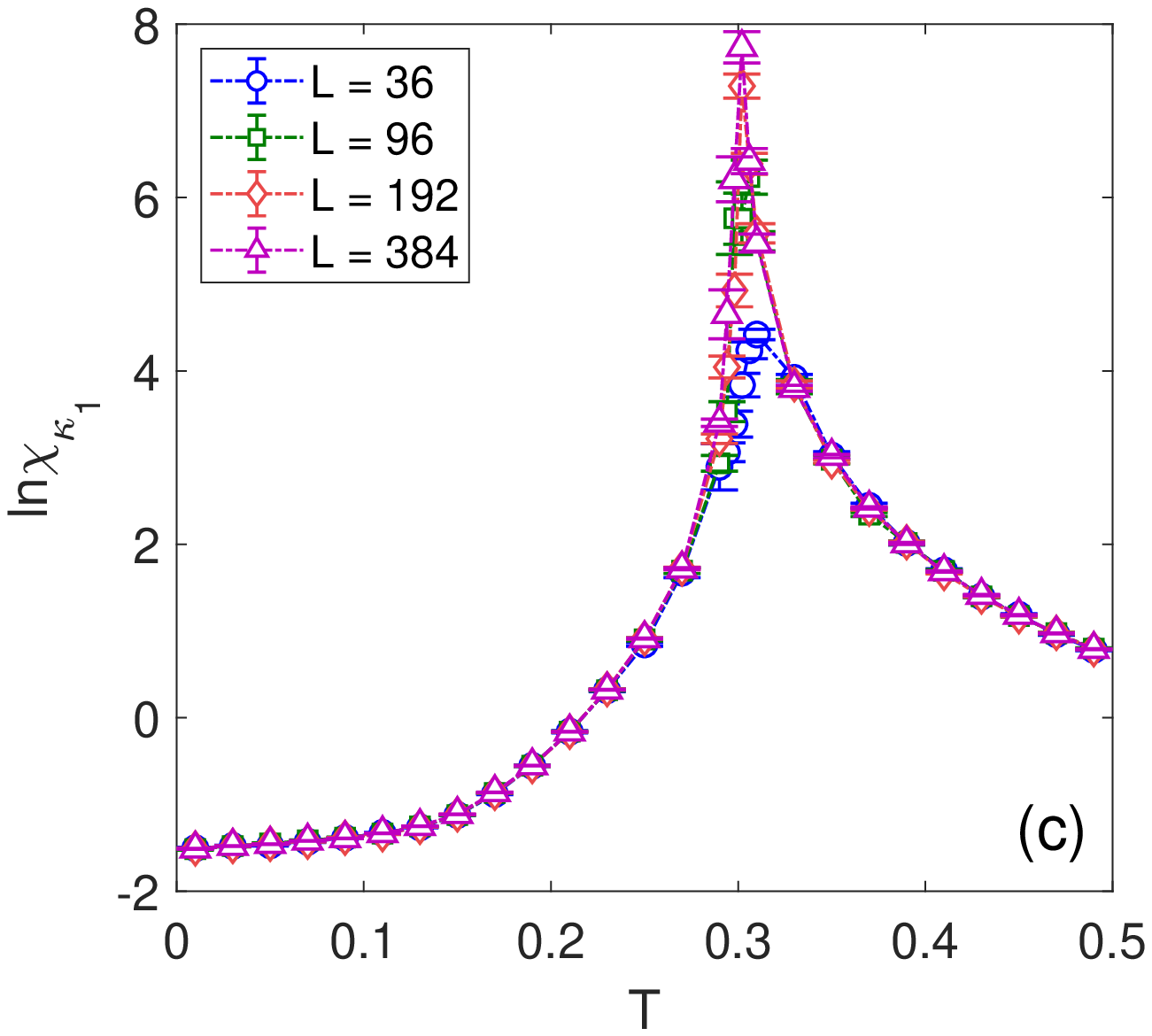}\label{fig:susc_chi1x08}}\\
\subfigure{\includegraphics[scale=0.4,clip]{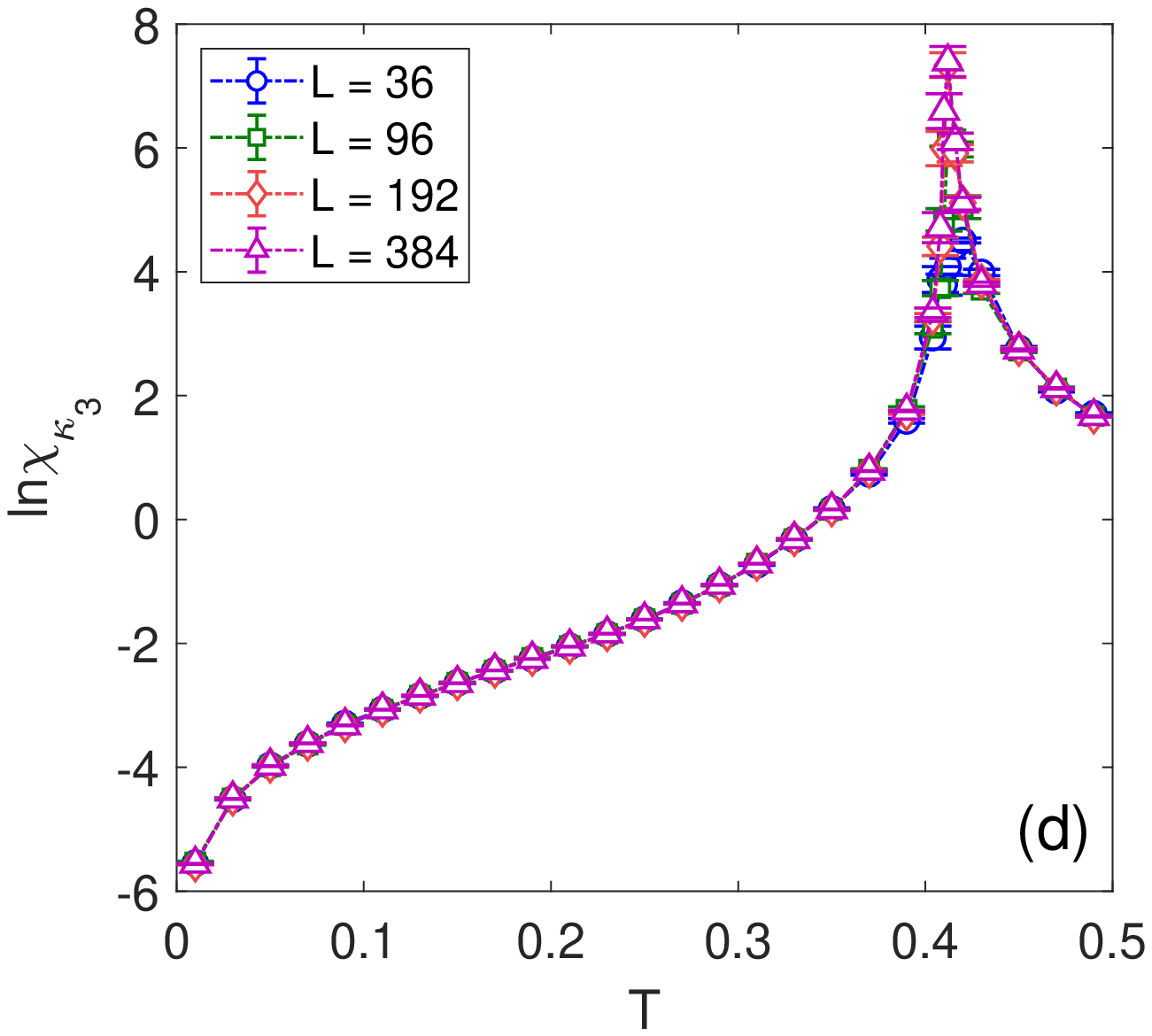}\label{fig:susc_chi3x02}}
\subfigure{\includegraphics[scale=0.4,clip]{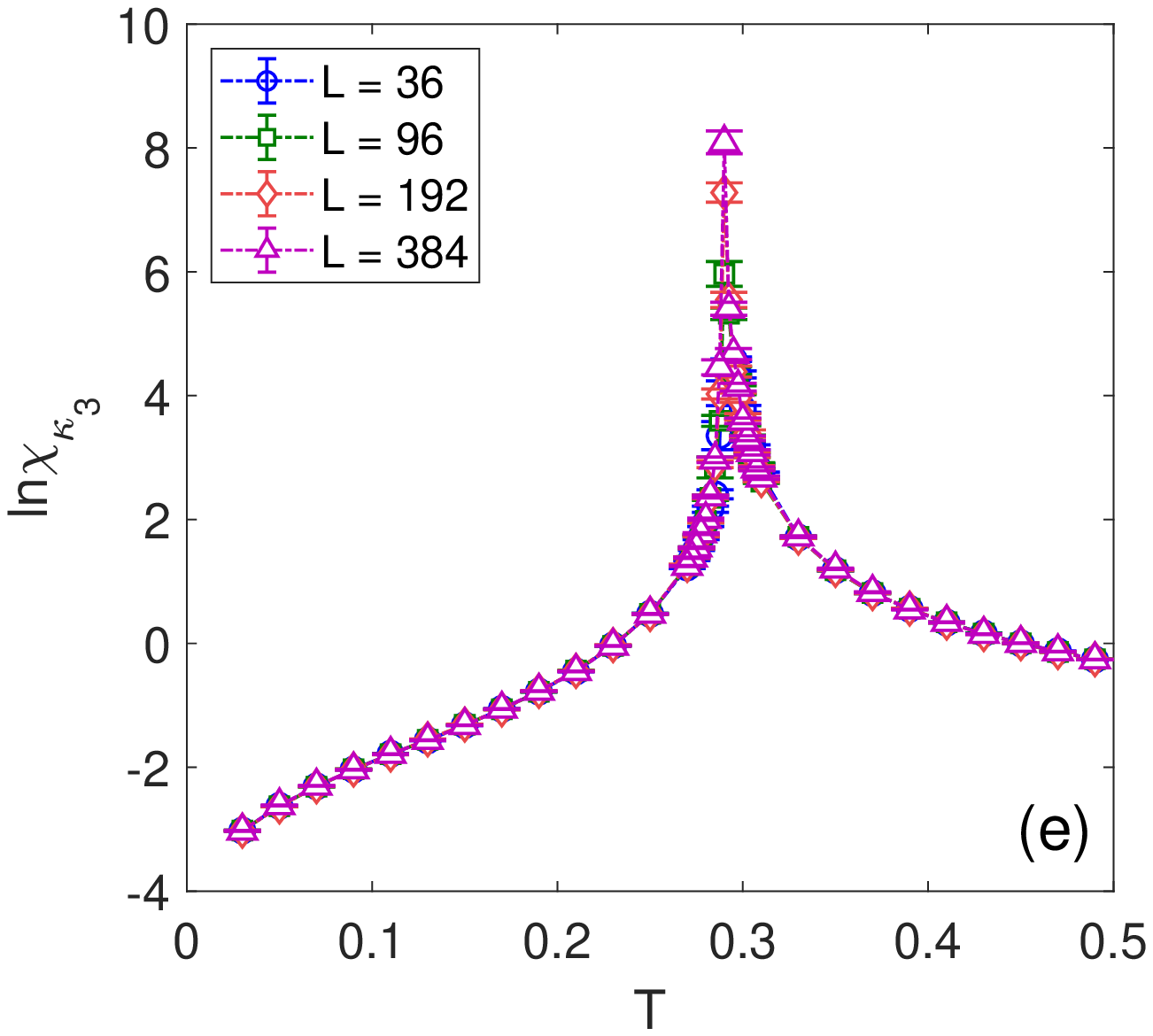}\label{fig:susc_kp3x0555}}
\subfigure{\includegraphics[scale=0.4,clip]{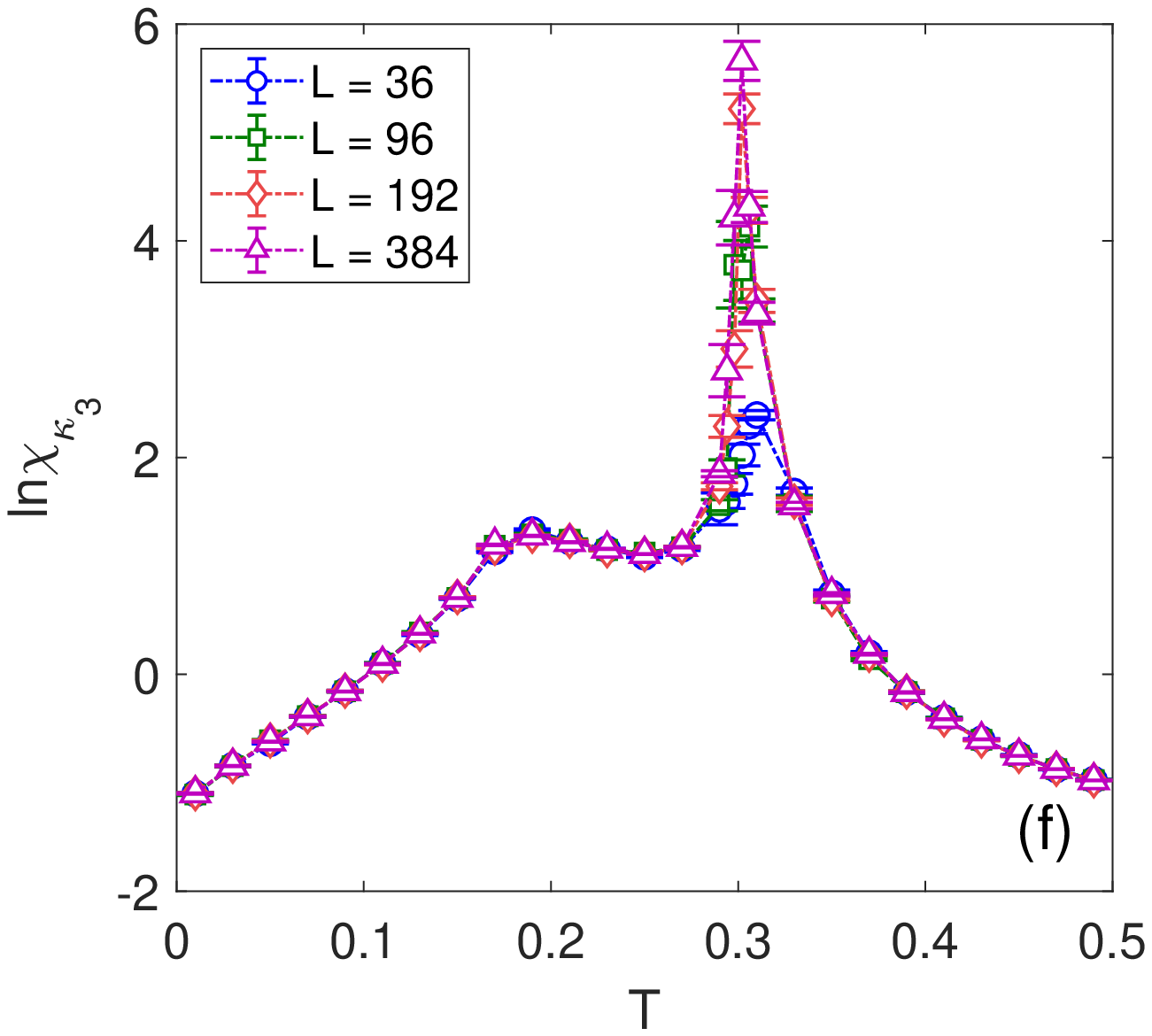}\label{fig:susc_chi3x08}}
\caption{Temperature dependencies of the generalized (staggered) chiral susceptibilities $\chi_{\kappa_1}$ (upper row) and $\chi_{\kappa_3}$ (lower row), for $\Delta=0.2$ (first column), $\Delta=0.555$ (second column), and $\Delta=0.8$ (third column), and different sizes $L$.}
\label{fig:chiral_susc}
\end{figure}

Observing the order parameters presented above for different lattice sizes, one can notice apparent size dependence in the quantities $m_1$ and $m_3$ in the whole temperature interval. On the other hand, the chiralities $\kappa_1$ and $\kappa_3$ only show noticeable dependence in the vicinity of the phase transition and within the paramagnetic phase. This points to different types of ordering of the respective quantities; while the vanishing of the former indicates QLRO the non-zero constant values of the latter signal true LRO. Based on the behavior of the previously studied models for $q=1$ and $q=2$, this scenario can be expected and it is also corroborated by the behavior of the respective susceptibilities, presented in Figs.~\ref{fig:magnetic_susc} and~\ref{fig:chiral_susc} on a semi-logarithmic scale. In the whole temperature intervals below transition temperatures, the magnetic and generalized nematic susceptibilities appear to diverge as power law, confirming the QLRO nature of the orderings. On the other hand, the chiral susceptibilities only diverge at the transition to the paramagnetic phase. Notice that at the low-temperature AN3-CAFM and AFM-CAFM transitions there are only round maxima in $\chi_{\kappa_1}$ (Fig.~\ref{fig:susc_chi1x02}) and $\chi_{\kappa_3}$ (Fig.~\ref{fig:susc_chi3x08}), respectively, insensitive to lattice size, and thus not related to any phase transitions.

\begin{figure}[t!]
\centering
\subfigure{\includegraphics[scale=0.4,clip]{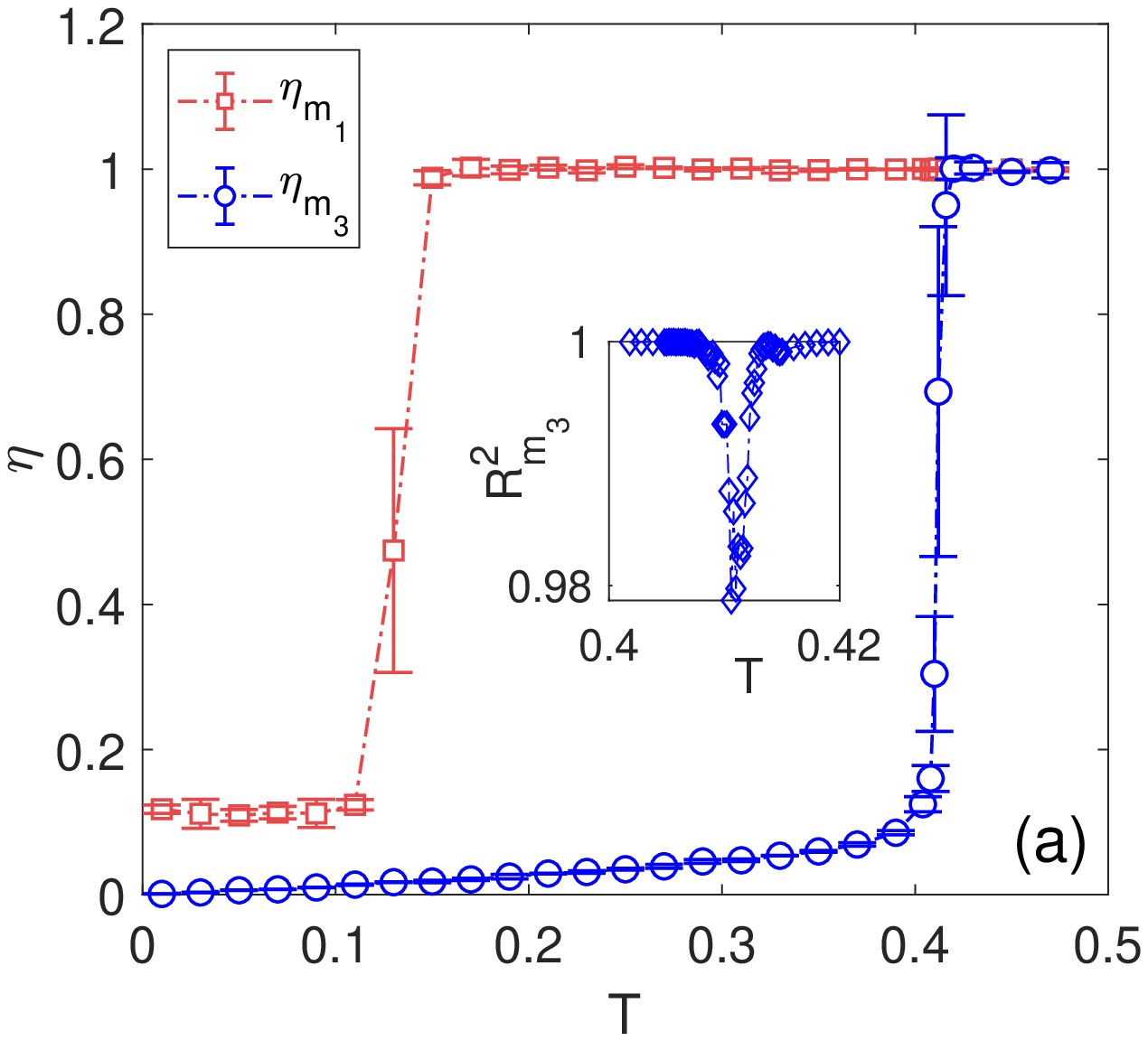}\label{fig:eta_m_02}}
\subfigure{\includegraphics[scale=0.4,clip]{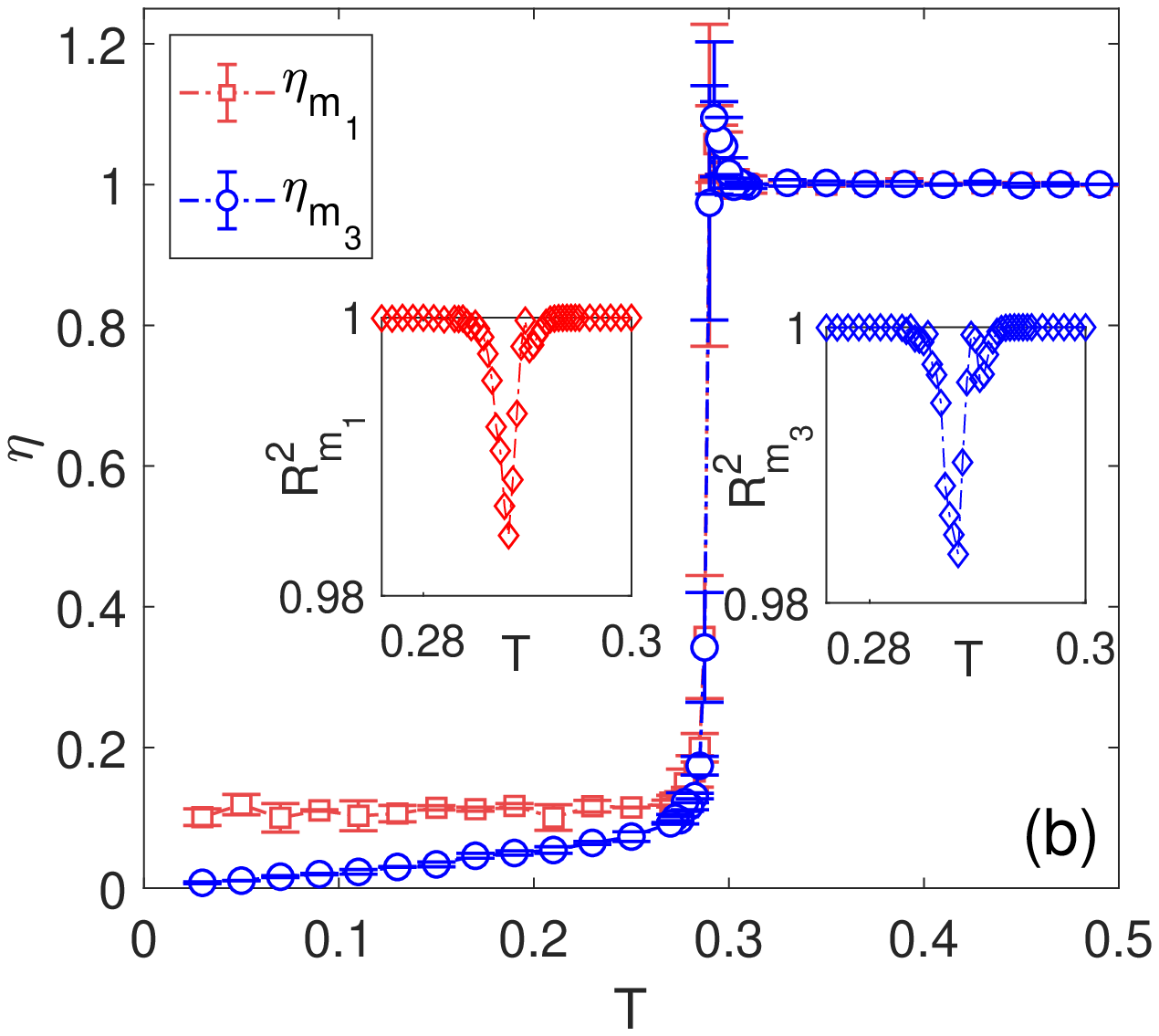}\label{fig:eta_m_0555}}
\subfigure{\includegraphics[scale=0.4,clip]{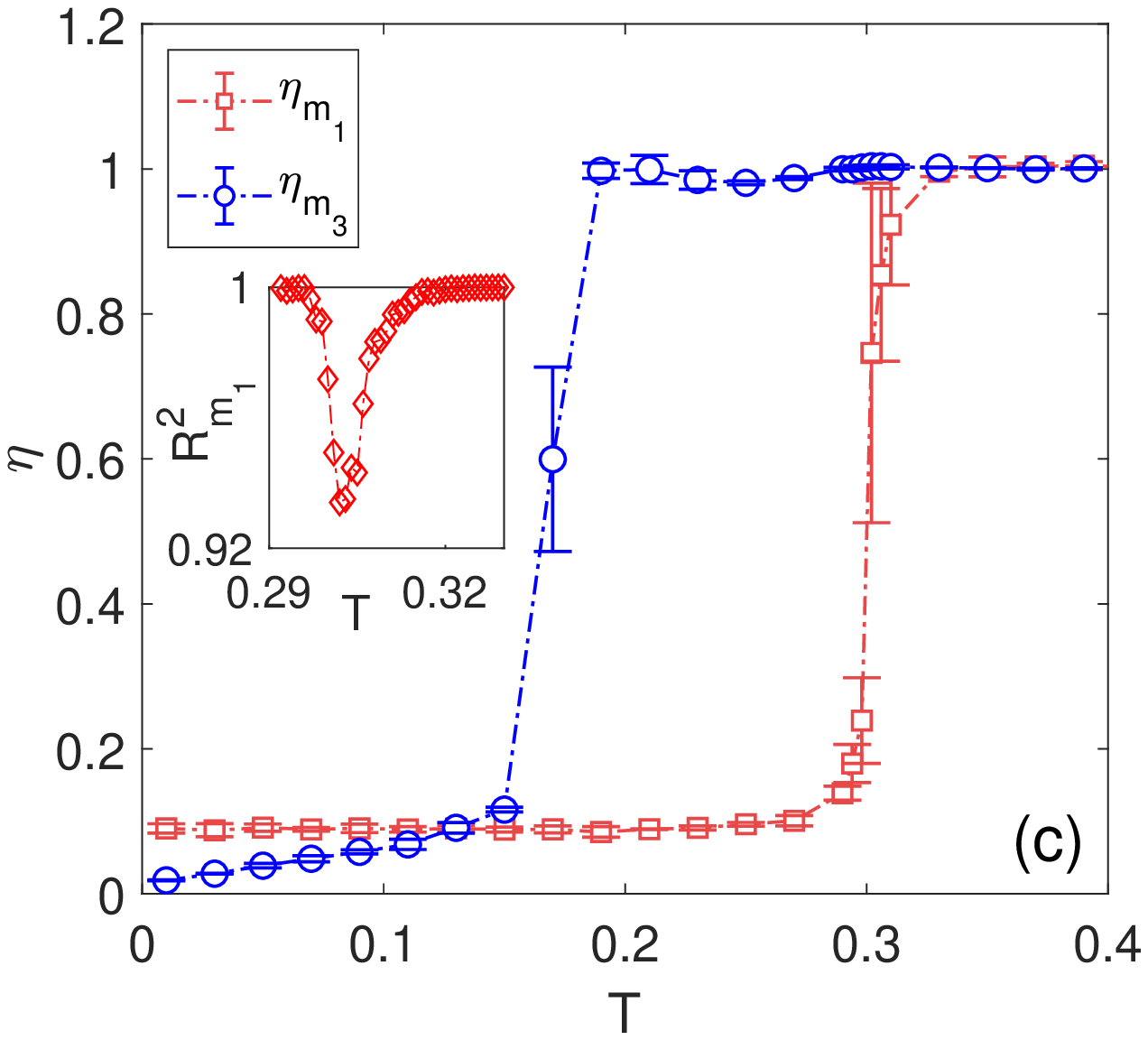}\label{fig:eta_m_08}}
\subfigure{\includegraphics[scale=0.4,clip]{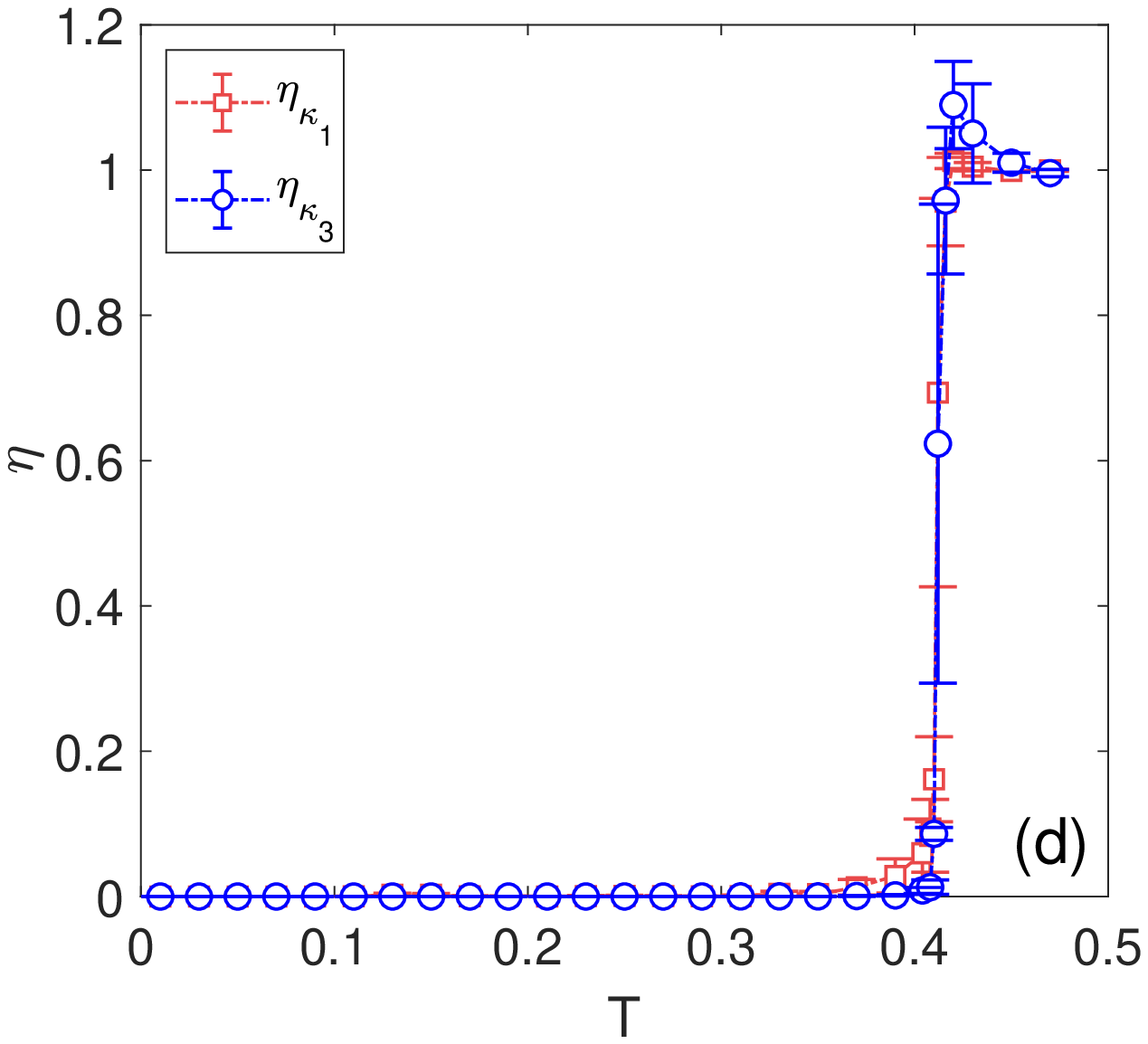}\label{fig:eta_kappa_02}}
\subfigure{\includegraphics[scale=0.4,clip]{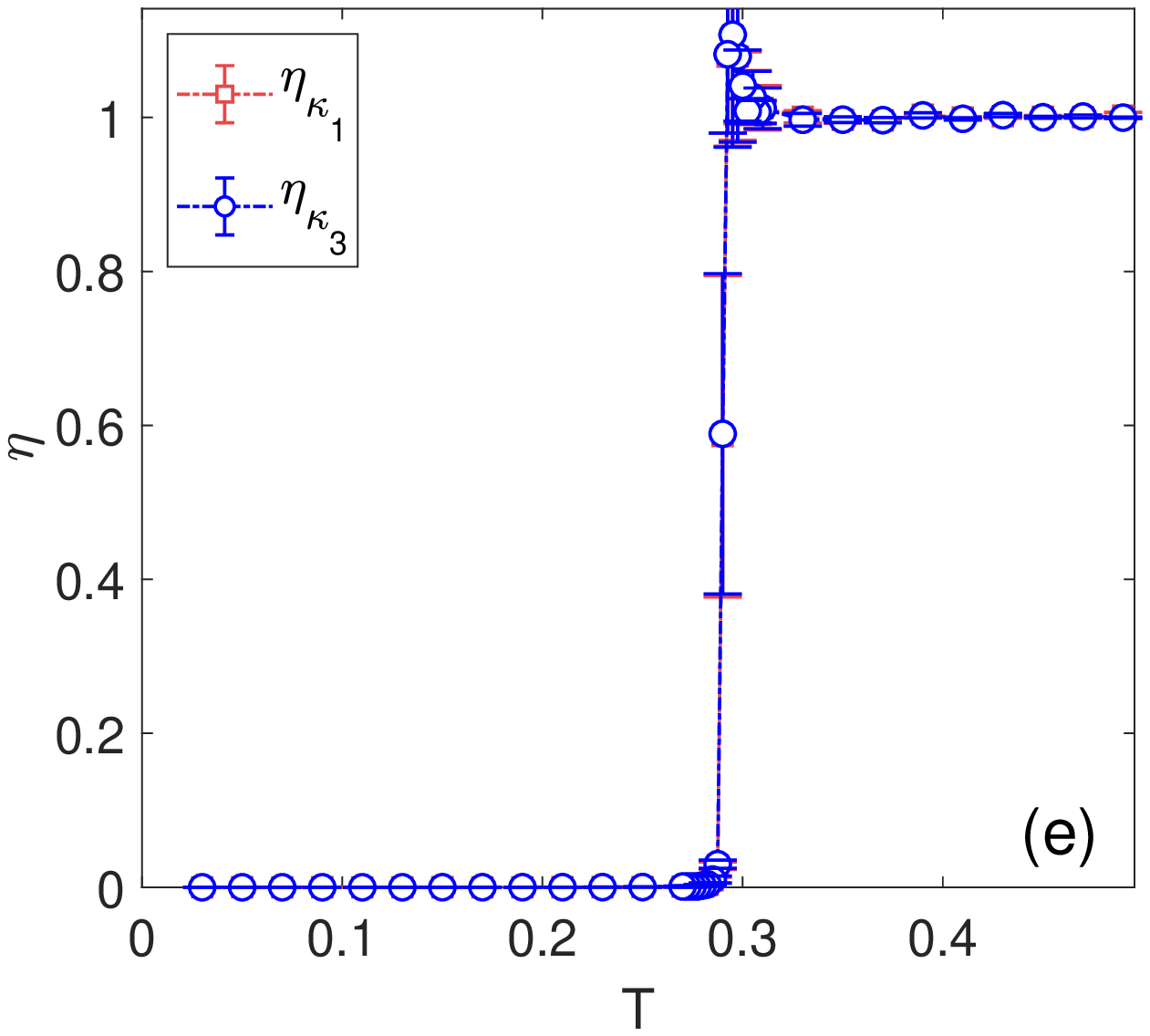}\label{fig:eta_kappa_0555}}
\subfigure{\includegraphics[scale=0.4,clip]{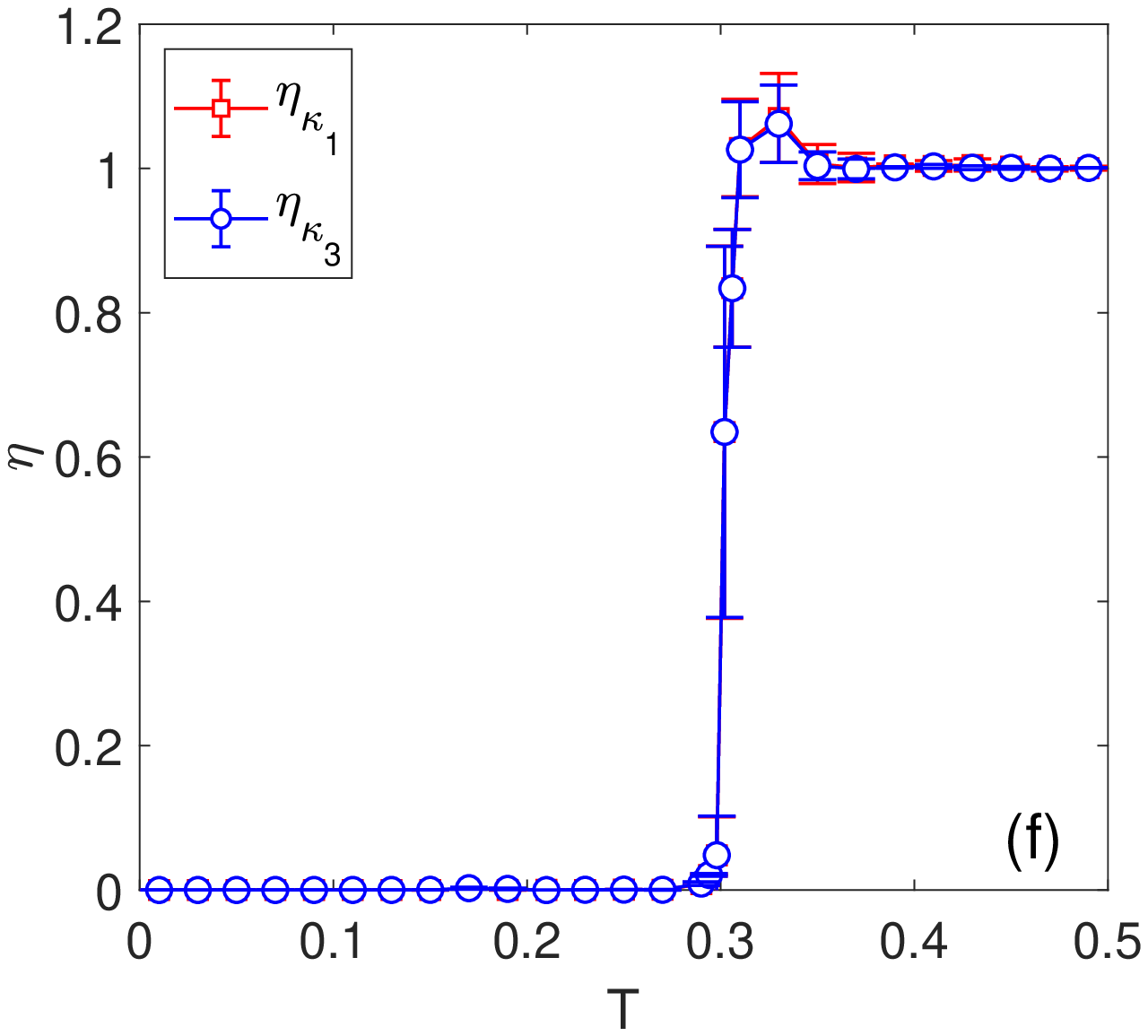}\label{fig:eta_kappa_08}}
\caption{The critical exponent $\eta$ for the magnetic and generalized nematic (upper row) and chiral and generalized chiral (lower row) order parameters, for $\Delta=0.2$ (first column), $\Delta=0.555$ (second column), and $\Delta=0.8$ (third column). The insets in the upper panels show the adjusted coefficient of determination $R^2$ in the vicinity of order-disorder transitions.}
\label{fig:critical_indices}
\end{figure}

Both the qualitative and quantitative characters of the decay of the order parameters $m_1$ and $m_3$ can be elucidated by performing their FSS analysis according to Eq.~(\ref{eq:fss_o_bkt}) in the whole temperature interval. By fitting the dependence of the order parameters on system size on a log-log scale, we obtain the temperature dependence of $\eta_o(T)$ depicted in Fig.~\ref{fig:critical_indices}. The values correspond to the negative of the slopes of the linear fits. At low temperatures, the values of the critical exponent $0 < \eta < 1$ for the magnetic and nematic order parameters (upper row) confirm the algebraic nature of the QLRO phases, while the jump to $\eta = 1$ signals the loss of the QLRO and the onset of the exponential decay of the correlation function, typical for the BKT phase transition. The order-disorder transition is partially smeared out by finite-size effects and, therefore, in order to determine the transition temperature more precisely it is useful to monitor the quality of the fits. In particular, we evaluate the adjusted coefficient of determination $R^2$, which can signal deterioration of the linear fit at the crossover between the two regimes if its value noticeably drops below one, as can be witnessed in the insets of Fig.~\ref{fig:critical_indices}. The values of the critical exponent $\eta$ at the transition temperatures, determined by such correlation analysis (marked in Fig.~\ref{fig:PD_q3} by filled squares), are presented in Table~\ref{table:eta_nu}. It is worth noticing that they all correspond, within statistical errors, to the value of $\eta^{\rm BKT}=1/4$, expected at the BKT transition~\cite{BKT1, BKT2}.

\begin{table}[t!]
\caption{Critical exponents $\eta$ at the order-disorder transition line (filled squares in Fig.~\ref{fig:PD_q3}), and $1/\nu$ at the AN3-CAFM and AFM-CAFM transition lines (filled circles in Fig.~\ref{fig:PD_q3}).}
\label{table:eta_nu}
\centering
\small
\begin{tabular}{ccccccccccc}
 \hline\hline
$\Delta$ & 0.1 & 0.2 & 0.3 & 0.4 & 0.5 & 0.555 & 0.6 & 0.7 & 0.8 & 0.9 \\
 \hline
$\eta$ & $0.270(21)$ & $0.254(8)$ & $0.249(8)$ & $0.238(6)$ & $0.239(15)$ & $0.250(8)$ & $0.252(17)$ & $0.242(9)$ & $0.244(17)$ & $0.252(8)$\\
$1/\nu$ & $-$ & $0.412(12)$ & $0.433(10)$ & $0.364(23)$ & $0.409(27)$ & $-$ & $0.472(59)$ & $0.474(3)$ & $0.505(9)$ & $-$\\   
 \hline\hline
\end{tabular}
\end{table}

As one would expect from the finite-size behavior of the chiral order parameters (last row in Fig.~\ref{fig:order}), the value of $\eta_{\kappa_k},\ k=1,3,$ is equal to zero for any temperature up to the transition temperature to the paramagnetic state at which it jumps to $\eta=1$ (lower row of Fig.~\ref{fig:critical_indices}). The chiral transition temperatures are depicted in Fig.~\ref{fig:PD_q3} by the cyan diamond symbols. Their values are very close to the BKT transition temperatures but slightly higher. The difference is observed for all values of $\Delta$ but the most clearly visible for $\Delta \approx 0.555$, as shown in the inset of Fig.~\ref{fig:PD_q3}.

\begin{figure}[t!]
\centering
\subfigure{\includegraphics[scale=0.5,clip]{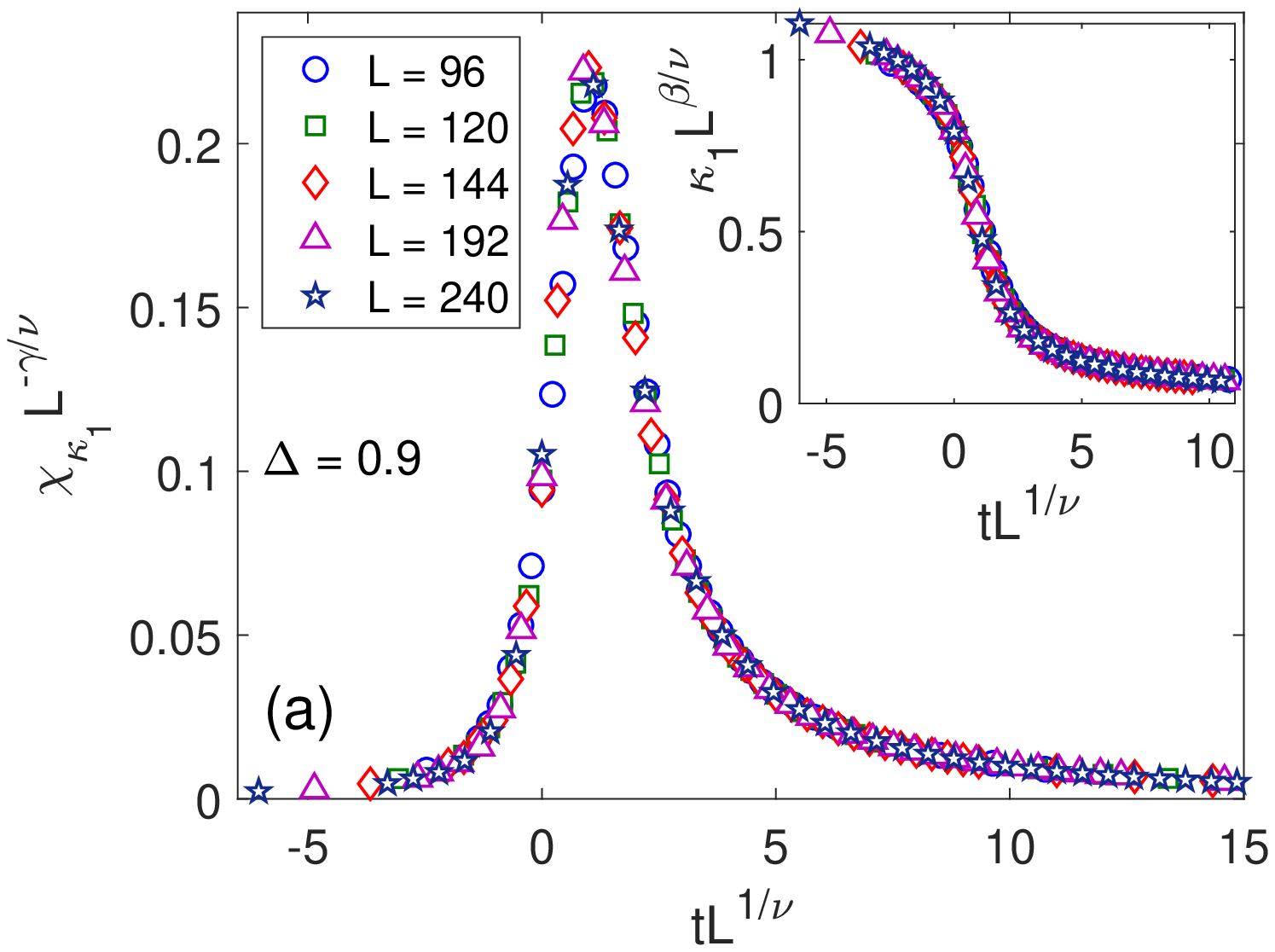}\label{fig:DataCollapse_kp1_x09_Comb}}
\subfigure{\includegraphics[scale=0.5,clip]{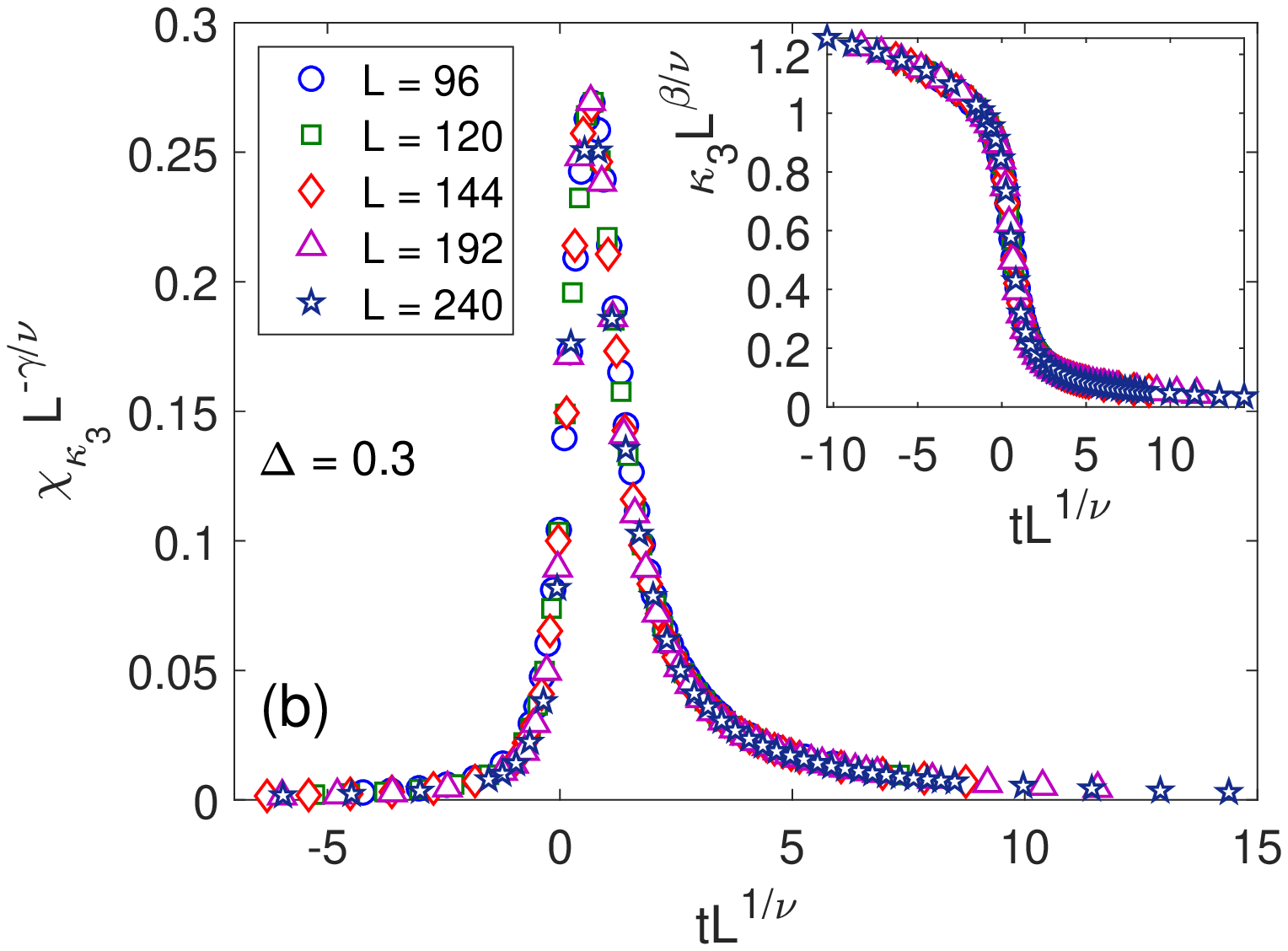}\label{fig:DataCollapse_kp3_x03_Comb}}
\subfigure{\includegraphics[scale=0.5,clip]{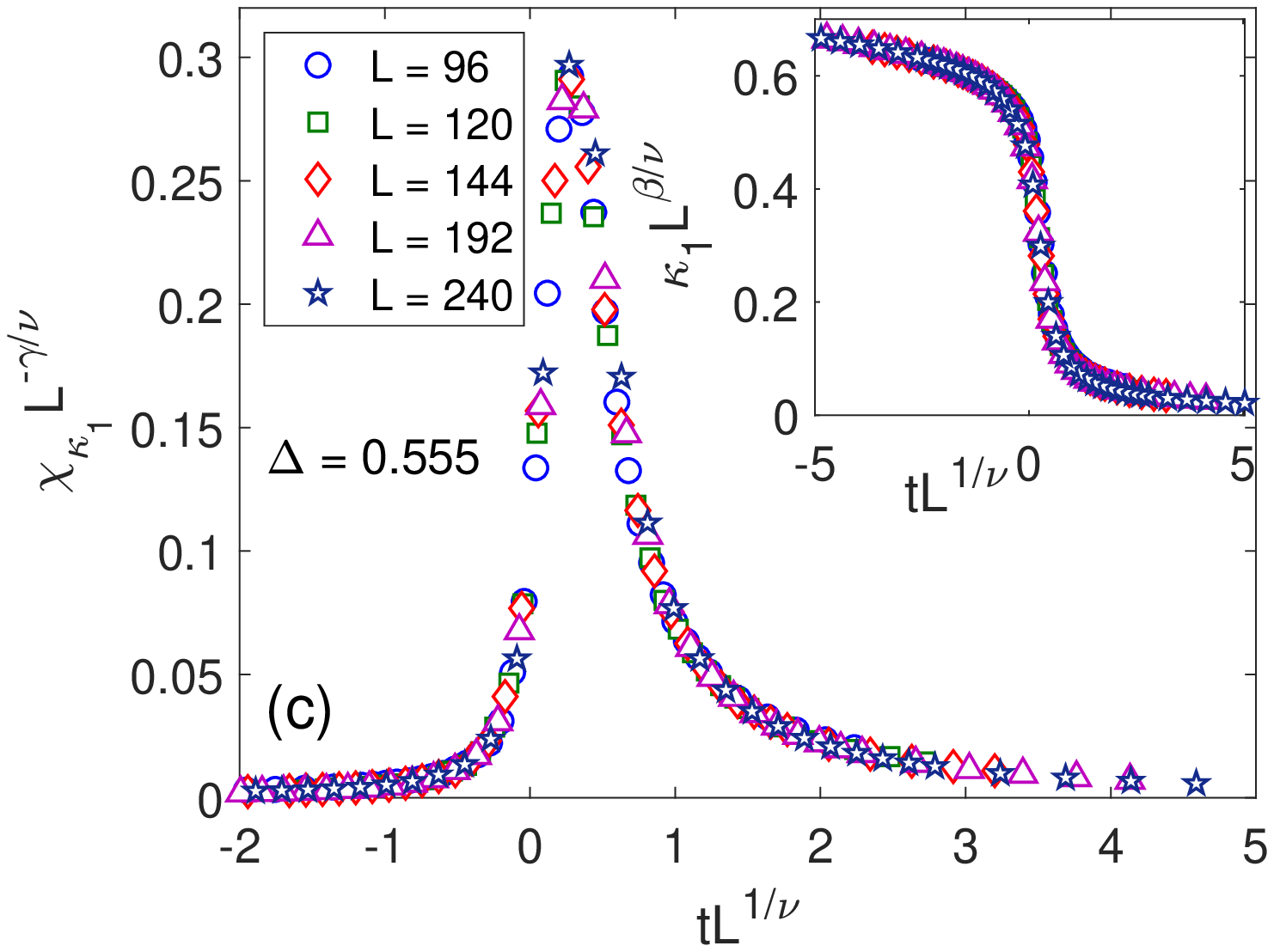}\label{fig:DataCollapse_kp1_x0555_Comb}}
\subfigure{\includegraphics[scale=0.5,clip]{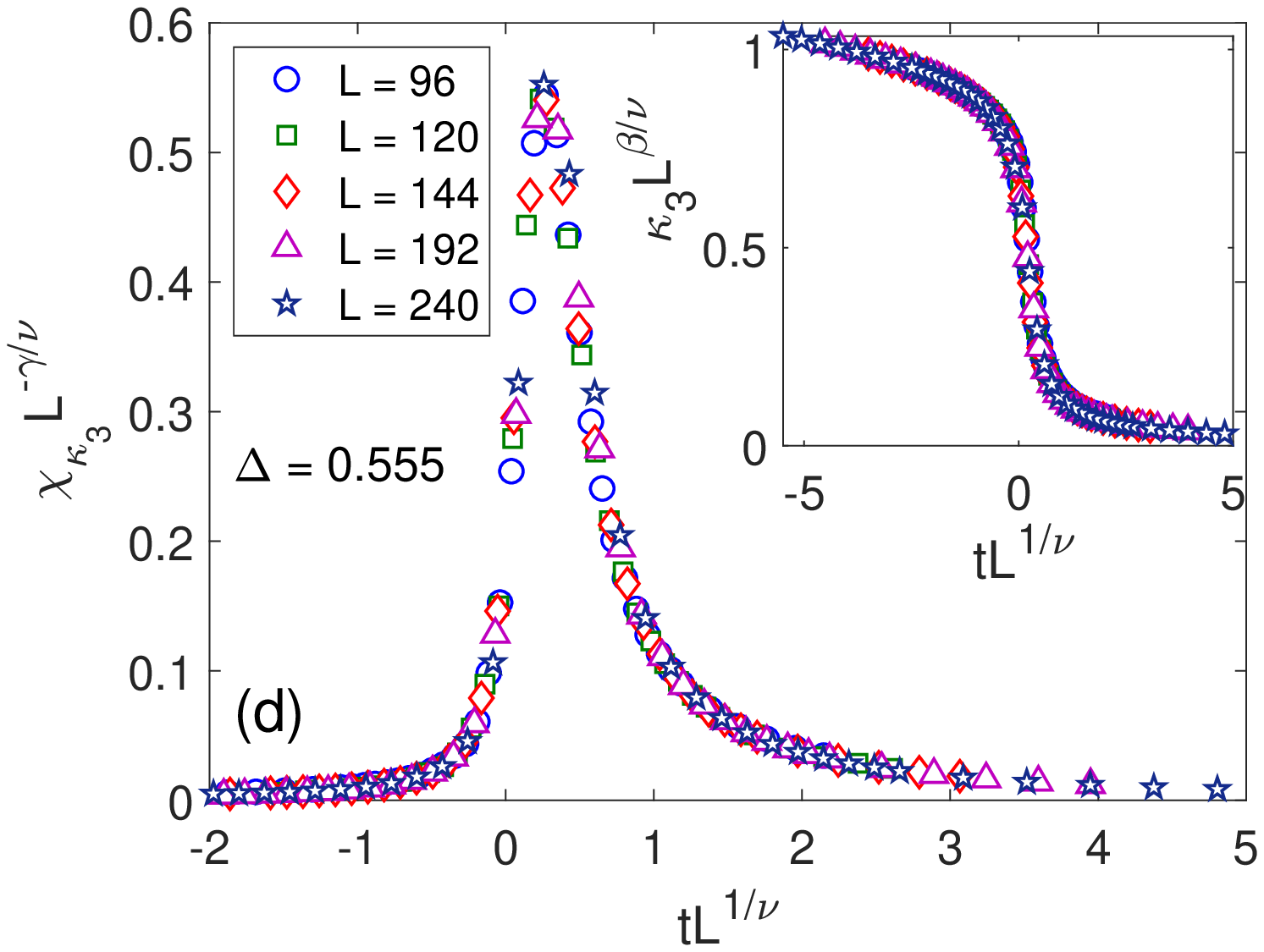}\label{fig:DataCollapse_kp3_x0555_Comb}}
\caption{Data collapse for chiral order parameters $\kappa_1$ and $\kappa_3$ (insets) and their susceptibilities $\chi_{\kappa_1}$ and $\chi_{\kappa_3}$ (main panels), for different values of $\Delta$ and the corresponding critical exponents listed in Table~\ref{table:chiralExp}.}
\label{fig:collapse_chir}
\end{figure}

The values of the critical exponents corresponding to the decay of the chiral order parameters, obtained from the data collapse of $\kappa_k$ and $\chi_{\kappa_k}, k=1,3$, at different values of $\Delta$~\footnote{For $\Delta=0.1$ and 0.2 the parameter $\kappa_1$ was too small for performing reliable data collapse analysis.}, are presented in Table~\ref{table:chiralExp}. For the frustrated $XY$ models on a triangular lattice with $q=1$ (standard $XY$ model)~\cite{Lee_1998} and the generalized model with $q=2$~\cite{Park-nematic} the chiral phase transition was concluded to be decoupled from the magnetic one with the critical exponents consistent with the three-state Potts model, $\nu_P=5/6$, $\gamma_P=13/9$ and $\beta_P=1/9$, albeit neither the possibility of the Ising universality class was ruled out. The chiral critical exponents of the present $q=3$ model apparently deviate from both the Ising as well as the three-state Potts universality classes. Interestingly, in contrast to the $q=1$ and $q=2$ models, the present values of the exponent $\nu$ are in a good agreement with the Ising value $\nu_I=1$. Also $\beta$ is fairly close to $\beta_I=1/8$ but $\gamma$ in most cases underestimate $\gamma_I=7/4$, expected for the Ising universality class. Nevertheless, the scaling relation $2\beta+\gamma=2\nu$ is fulfilled within the error bars for all values of $\Delta$, except in the vicinity of $\Delta=0.555$, where all the phase boundaries meet. The excellent data collapse of the chiralities and chiral susceptibilities with the non-Ising critical exponents, listed in Table~\ref{table:chiralExp}, is demonstrated in Fig.~\ref{fig:collapse_chir} for some representative values of $\Delta$.

\begin{table}[t!]
\caption{Critical exponents $\gamma, \nu$, and $\beta$, at the chirality $\kappa_1$ and $\kappa_3$ phase transitions.}
\label{table:chiralExp}
\centering
\begin{tabular}{p{2cm}p{1.6cm}p{1.6cm}p{1.6cm}p{1.6cm}p{1.6cm}p{1.6cm}}
 \hline\hline
  &\multicolumn{3}{c}{$\kappa_1$ transition} &   \multicolumn{3}{c}{$\kappa_3$ transition}\\
\hline
$\hfil\Delta$ & $\hfil\gamma$ & $\hfil\nu$ & $\hfil\beta$ & $\hfil\gamma$ & $\hfil\nu$ & $\hfil\beta$ \\
 \hline
\hfil0.1&   $\hfil -$&    $\hfil -$&   $\hfil - $&   $\hfil 1.7132$&    $\hfil 1.0046 $&    $\hfil 0.105 $\\
 \hfil0.2&   $\hfil -$&    $\hfil -$&   $\hfil - $&   $\hfil 1.6817$&    $\hfil 1.00 $&    $\hfil 0.107 $\\
 \hfil0.3&   $\hfil 1.5394$&    $\hfil 1.0047$&   $\hfil 0.150 $&   $\hfil 1.7220$&    $\hfil 1.0226 $&    $\hfil 0.115 $\\
 \hfil0.4&   $\hfil 1.7508$&    $\hfil 1.06$&   $\hfil 0.102 $&   $\hfil 1.8183$&    $\hfil 1.10 $&    $\hfil 0.098 $\\
 \hfil0.5&   $\hfil 1.6598$&    $\hfil 1.0671$&   $\hfil 0.125 $&   $\hfil 1.5567$&    $\hfil 1.0642 $&    $\hfil 0.125 $\\
\hfil 0.555&   $\hfil 1.7277$&    $\hfil 1.1261$&   $\hfil 0.084 $&   $\hfil 1.7462$&    $\hfil 1.1372 $&    $\hfil 0.100$\\
 \hfil0.6&   $\hfil 1.6514$&    $\hfil 1.00$&   $\hfil 0.1395 $&   $\hfil 1.6769$&    $\hfil 1.0117 $&    $\hfil 0.1403 $\\
 \hfil0.7&   $\hfil 1.6845$&    $\hfil 1.00$&   $\hfil 0.110 $&   $\hfil 1.7535$&    $\hfil 1.03 $&    $\hfil 0.115 $\\
 \hfil0.8&   $\hfil 1.6995$&    $\hfil 0.9615$&   $\hfil 0.125 $&   $\hfil 1.7416$&    $\hfil 0.9784 $&    $\hfil 0.125 $\\
\hfil0.9&   $\hfil 1.7098$&    $\hfil 1.0199$&   $\hfil 0.125 $&   $\hfil 1.6898$&    $\hfil 1.0199 $&    $\hfil 0.125 $\\
 \hline\hline
\end{tabular}
\end{table}

\subsubsection{Magnetic-nematic transitions}

\begin{figure}[t!]
\centering
\subfigure{\includegraphics[scale=0.37,clip]{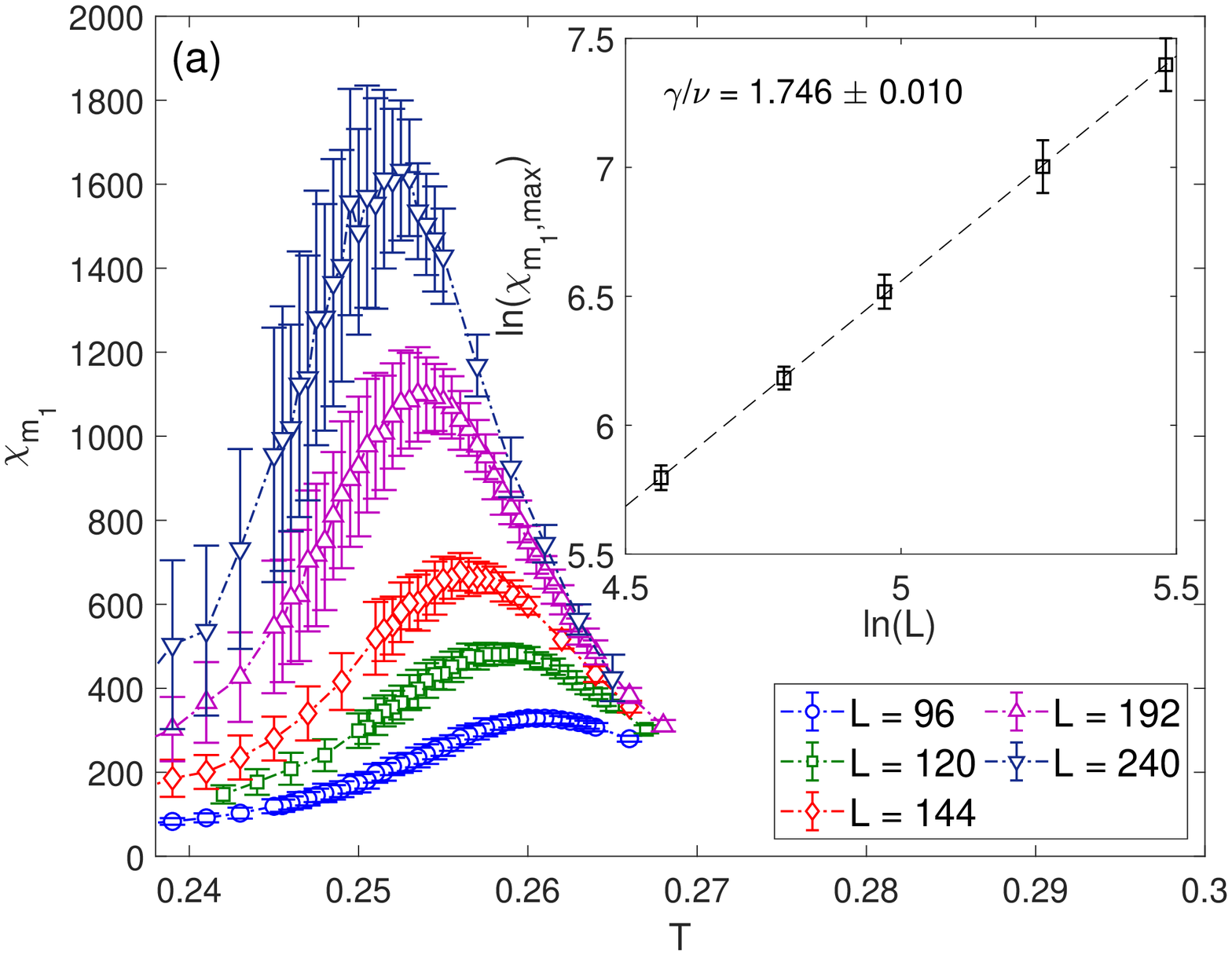}\label{fig:Scaling_x04}}
\subfigure{\includegraphics[scale=0.37,clip]{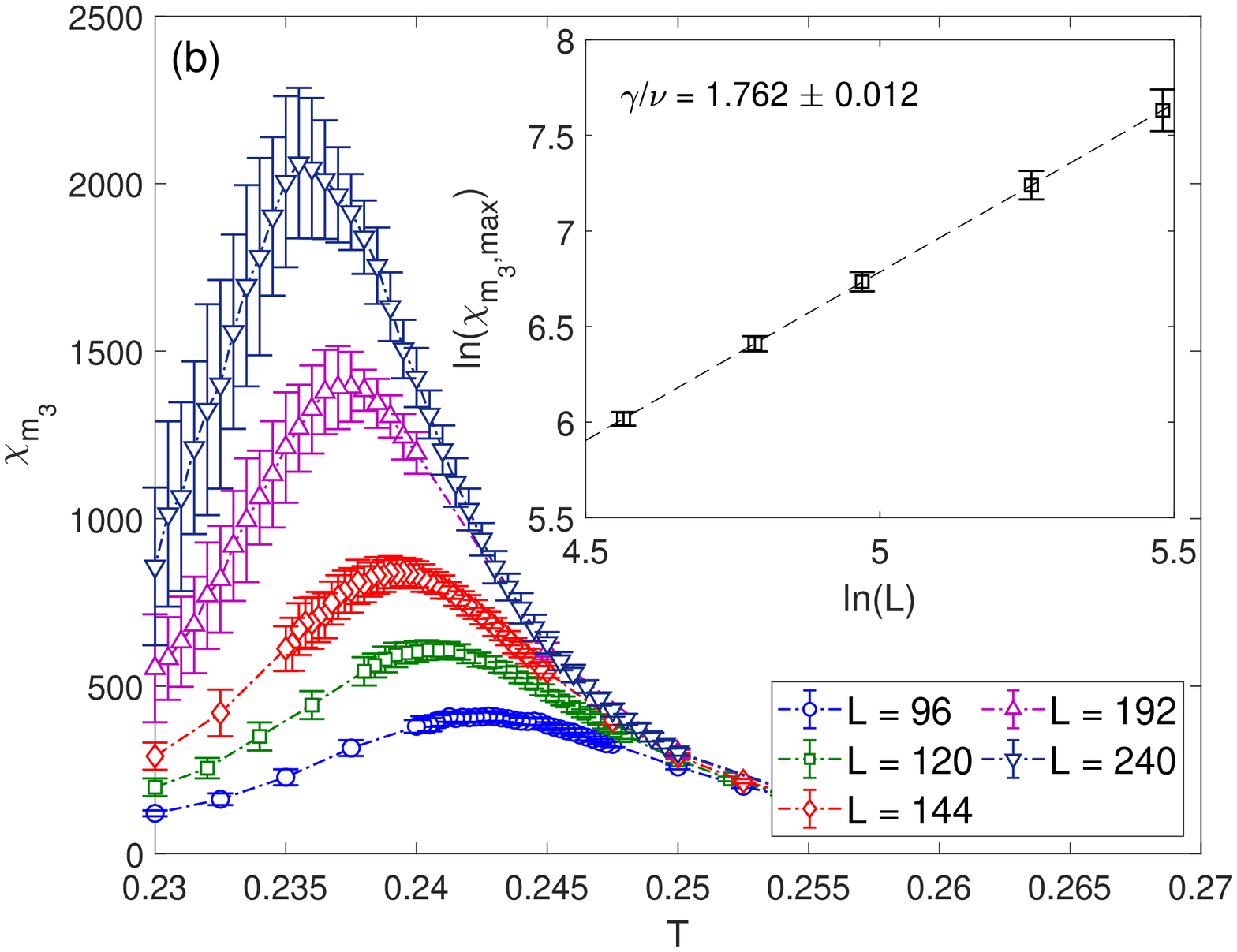}\label{fig:Scaling_x07}}\\
\subfigure{\includegraphics[scale=0.37,clip]{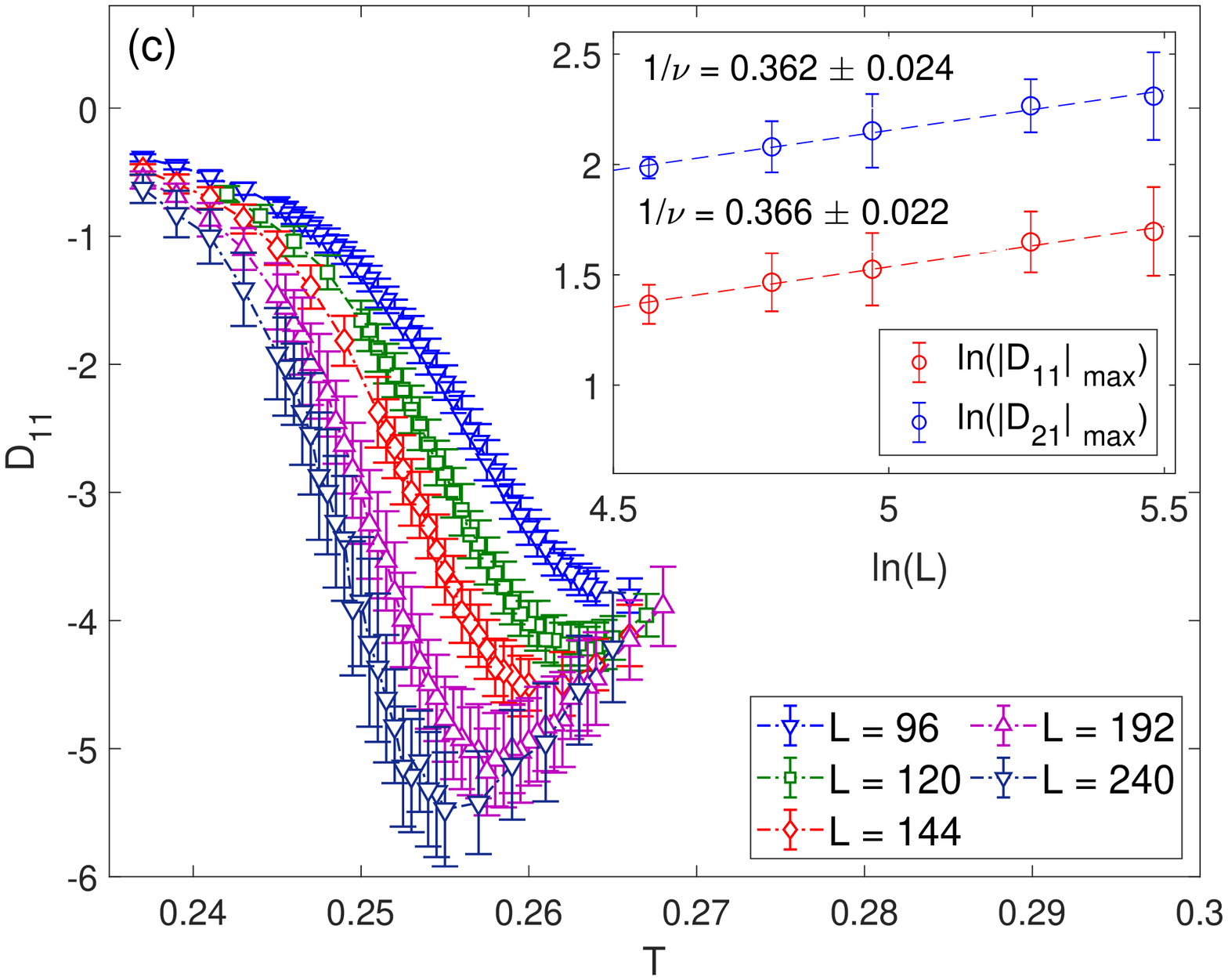}\label{fig:Scaling_x04D11}}
\subfigure{\includegraphics[scale=0.37,clip]{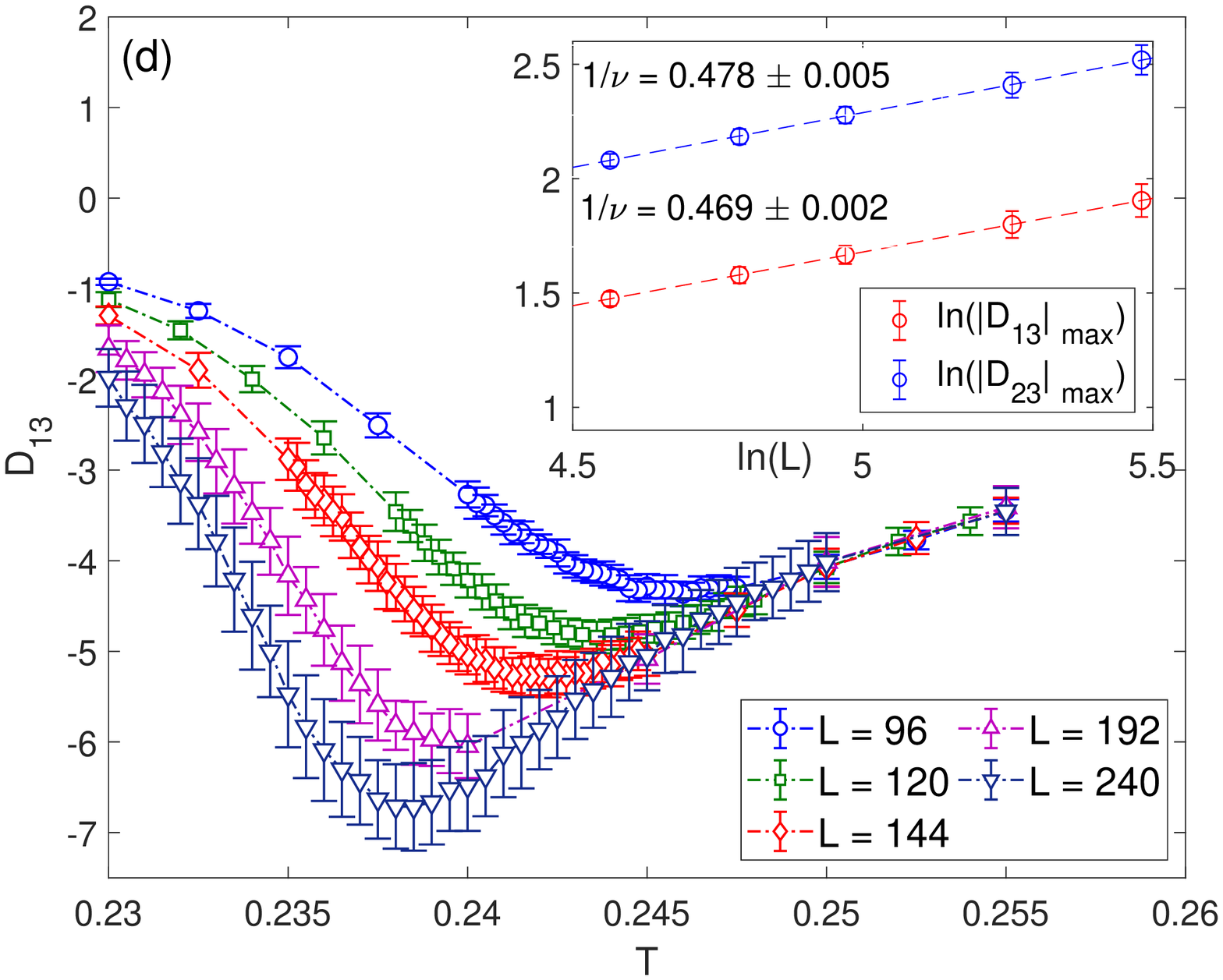}\label{fig:Scaling_x07D13}}
  \caption{Determination of the critical exponent ratios $\gamma/\nu$ (a,b) and $1/\nu$ (c,d), using the FSS method at $\Delta=0.4$ (a,c) and $\Delta=0.7$ (b,d).}
  \label{fig:susc_scaling}
\end{figure}

Let us now focus on the character of the phase transitions between the identified QLRO phases, i.e., the AN3-CAFM and AFM-CAFM transitions, which occur at lower temperatures within $0 < \Delta  \lesssim 0.997$. To study the critical behavior in this region for each lattice size, we ran $100$ independent MC simulations with up to $1.6\times10^7$ sweeps per temperature step, to obtain temperature dependencies of the mean values of the quantities~(\ref{eq:susc}) and~(\ref{eq:log_der}) by configurational averaging. Those were subsequently used in the FSS analysis to obtain the critical exponents ratios. The results for selected values of $\Delta=0.4$ and 0.7, representing the two branches of the phase boundaries, are presented in Fig.~\ref{fig:susc_scaling}. One can notice that the error bars considerably increase with lattice size, which can be attributed to the gradual increase of autocorrelation times as one goes to still lower temperatures and larger system sizes. We note that the results of FSS analysis for $\Delta=0.1$ and 0.9, with the transition points located at very low temperatures, are not included due to problems related to very large autocorrelation times reaching the order of $10^5$ for the largest $L$. 

Nevertheless, we were able to determine the critical exponents ratios with reasonably high precision (see the insets). The values of $\gamma/\nu = 1.746 \pm 0.010$ for $\Delta=0.4$ (AN3-CAFM transition) and $\gamma/\nu = 1.762 \pm 0.012$ for $\Delta=0.7$ (AFM-CAFM transition) are close to the Ising values, but in the former case the three-state Potts universality cannot be ruled out either. However, the corresponding values of $1/\nu$ for both transitions are, beyond any doubt, different from either universality class and their values change with $\Delta$ (see Table~\ref{table:eta_nu}). Consequently, also the values of $\gamma$ will not be compatible with any universality class. On the other hand, the ratios of $\gamma/\nu$ along both transition boundaries appear to be constant and compatible within the error bars with both Ising and three-state Potts values. This points to the possibility of the \emph{weakly} universal behavior~\cite{suzuki_WU} governed by either the Ising or three-state Potts critical exponents ratios.

\begin{figure}[t!]
\centering
\subfigure{\includegraphics[scale=0.55,clip]{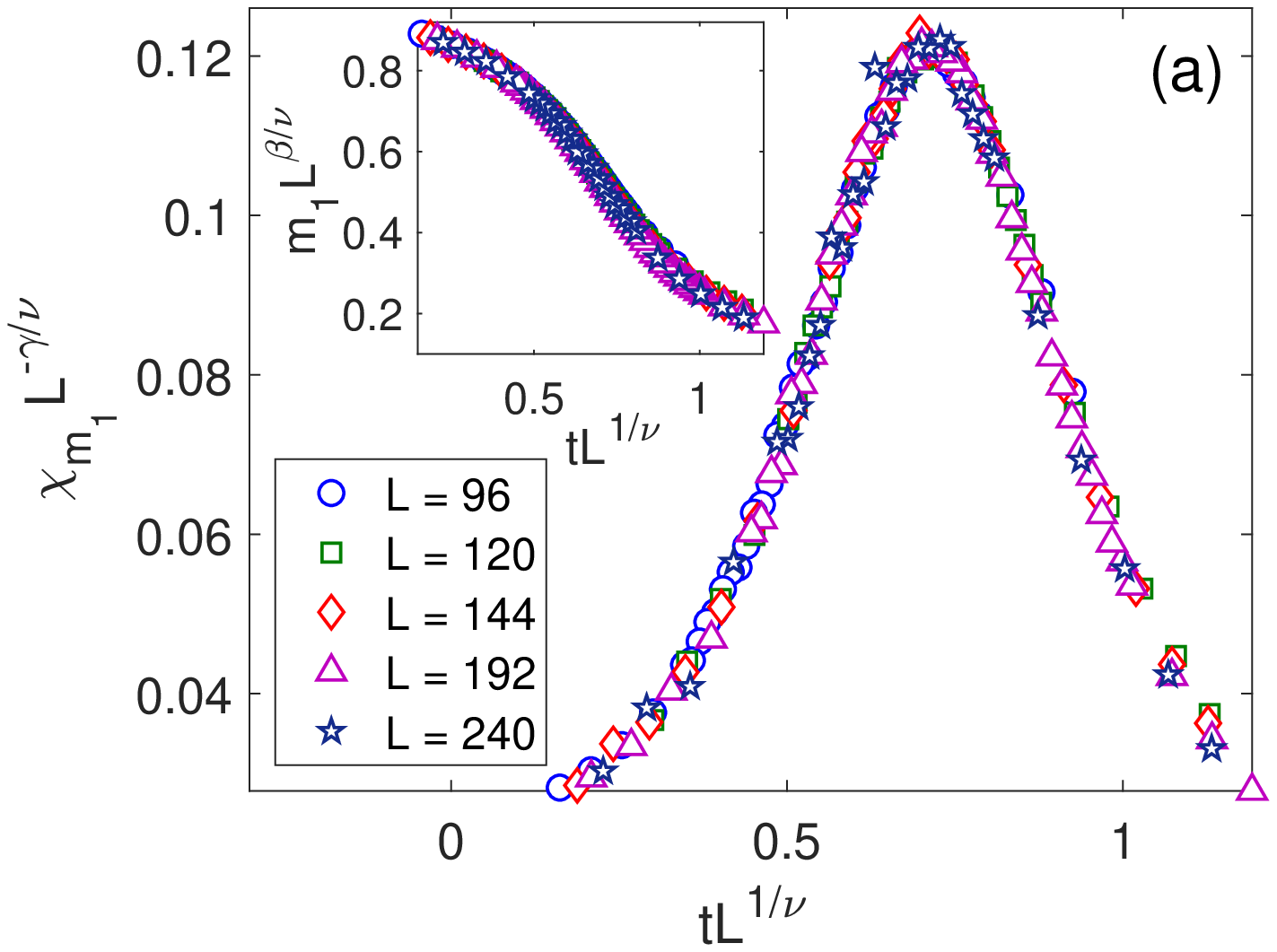}\label{fig:DataCollapse_m1Susc_x04_Weak_Potts_comb}}
\subfigure{\includegraphics[scale=0.55,clip]{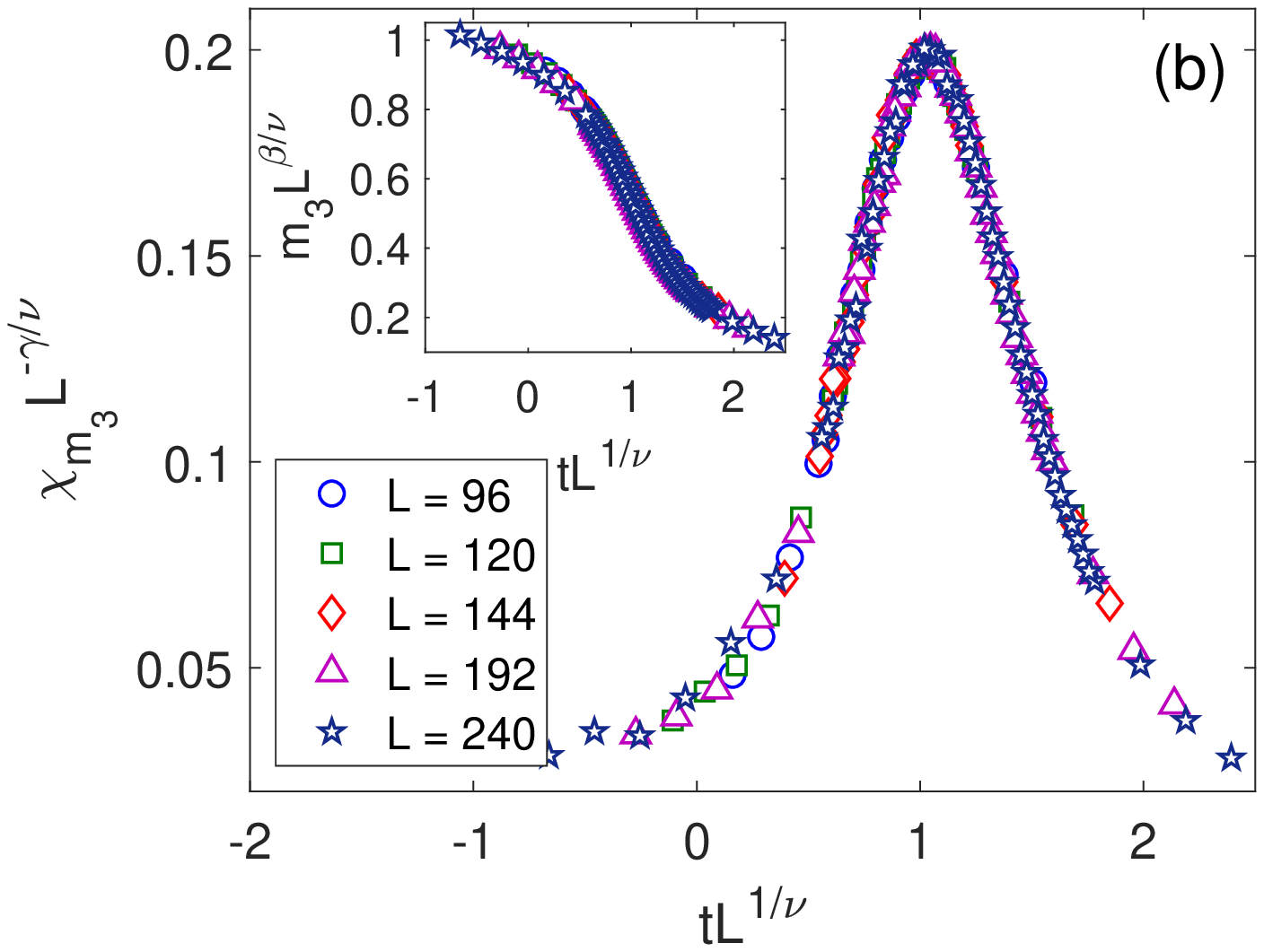}\label{fig:DataCollapse_m3Susc_x08_Weak_Ising_comb}}
\caption{Data collapse of (a) the magnetic susceptibility $\chi_{m_1}$ and order parameter $m_1$ (inset) for $\Delta = 0.4$ using the three-state Potts critical exponents ratios, and (b) the nematic susceptibility $\chi_{m_3}$ and order parameter $m_3$ (inset) for $\Delta = 0.8$ using the Ising critical exponents ratios.}
\label{fig:collapse_magn}
\end{figure}

To determine the proper weak universality class along the low-temperature phase transition boundaries we performed the data collapse analysis of the respective order parameters and the corresponding susceptibilities, using both the Ising and the three-state Potts values. We found that a noticeably better collapse could be obtained using the three-state Potts values at the AN3-CAFM boundary, i.e. for $0.2 \leq \Delta < 0.5$, while the Ising values gave better results along the AFM-CAFM boundary, i.e. for $0.6 < \Delta \leq 0.8$. The results for $\Delta=0.4$ and $\Delta=0.8$ are presented in Fig.~\ref{fig:collapse_magn}. Within $0.5\leq \Delta \leq 0.6$ the critical behavior is affected by proximity of different phase boundaries. We were not able to achieve a reasonably good collapse of neither the order parameters nor the susceptibilities. This may signify a direct BKT-type transition from the paramagnetic to the CAFM phase, as already suggested by the behavior of different thermodynamic quantities presented in Fig.~\ref{fig:order}.

The corresponding transition temperatures estimated both roughly from the specific heat maxima (empty circles), as well as more precisely from the data collapse analysis (filled circles), are shown in Fig.~\ref{fig:PD_q3}. It is interesting to notice that the former appreciably overestimate the true values along the AN3-CAFM and AFM-CAFM branches (especially at higher temperatures), while the values of the BKT transition temperatures estimated from the specific heat peaks coincide rather well with those from the FSS analysis. This is in contrast with the nonfrustrated generalized $XY$ model on a square lattice~\cite{Hubscher_2013}, were the opposite phenomenon has been observed. Consequently, the more precise location of the low-temperature branches of the phase diagram from the data collapse analysis provides evidence that in fact they do not connect to the high-temperature BKT branch, as the specific heat maxima would suggest, but rather only touch it at $\Delta \approx 0.555$.

\section{Summary and discussion}
For the first time we studied critical behavior of the generalized $XY$ model on a triangular lattice with $q = 3$, which includes both geometrical frustration as well as competition between the antiferromagnetic (AFM) and generalized antinematic (AN3) interactions. It has been previously shown that inclusion of even very small AN3 interaction changes the ground state from the 120-degree AFM structure to a peculiar canted (CAFM) state~\cite{zukovic_frustrated_2016}. In the present study we demonstrated that at finite temperatures the model features three phases: two high-temperature phases with purely AFM and purely AN3 types of ordering and one low-temperature CAFM phase, wedged between the AFM and AN3 phases, with mixed AFM and AN3 type of quasi-long-range ordering (QLRO). Thus, for almost any interaction strength ratio there are two phase transitions: the high-temperature order-disorder transition to either AFM or AN3 phase, followed by another transition at lower temperatures to the CAFM phase. Two exceptions include the limit of very small AN3 interaction, where only the AFM phase is present, and a very narrow region of comparable AFM and AN3 interactions, where only the CAFM phase seems to be present with a direct transition to the paramagnetic state. While the high-temperature order-disorder transitions are of the BKT type, the low-temperature transitions between the AFM-CAFM and AN3-CAFM phases are concluded to belong to the 2D weak Ising and weak three-state Potts universality classes, respectively. Besides the magnetic and nematic QLRO, all the identified phases also feature true LRO of the standard and generalized chiralities. They both vanish simultaneously at the second-order chiral phase transitions with the critical exponents deviating from the Ising universality and the critical temperatures slightly higher than the magnetic and nematic BKT transition temperatures. 

It is interesting to compare the present results with those obtained for the geometrically frustrated generalized model on a triangular lattice with $q=2$~\cite{Park-nematic} as well as the nonfrustrated model for $q = 3$ on a square lattice with ferromagnetic (FM) and generalized nematic (N3) interactions~\cite{Poderoso-2011, Canova-2014, Canova-2016}. While both these models display similar phase diagrams with separate magnetic and nematic phases, the topology of the present phase diagram is different. Namely, it features an additional new (CAFM) phase, which results from the competition between the two couplings absent in the above models. It is characterized by the coexisting AFM and AN3 QLRO as well as the chiral LRO. On the other hand, the high-temperature AFM (AN3) phase still coexists with the chiral LRO but lacks AN3 (AFM) QLRO. In the nonfrustrated $q = 3$ model on a square lattice, the nematic-magnetic N3-FM phase transition was found to belong to the three-state Potts universality class. In the present frustrated model the critical exponents at the corresponding AN3-CAFM transition do not comply with the three-state Potts universality class but their ratios $\beta/\nu$ and $\gamma/\nu$ do, i.e., the weak universality is valid. On the other hand, the new AFM-CAFM phase boundary appears to belong to the weak Ising universality class. The Ising-like character of this transition can be related to selecting one of the angles $\pm 3\pi/2$ from two associated canted states when crossing from CAFM to AFM phases (see Ref.~\cite{zukovic_frustrated_2016}). Finally, similar to the frustrated $q = 2$ model on a triangular lattice, also for the $q = 3$ case we found evidence of decoupling of magnetic/nematic and chiral phase transitions, albeit the critical exponents of the latter were different from the $q = 2$ case.   

The above findings raise further questions regarding the relevance of the higher-order couplings for the critical behavior of the continuous $XY$ models. For the nonfrustrated generalized $XY$ model on a square lattice it was found that for $q \geq 5$, the topology of the phase diagram changes and new phases emerge~\cite{Poderoso-2011,Canova-2016}. In the present frustrated model a new phase appeared already for $q = 3$. It would be interesting to extend the present investigation to include higher-order terms ($q > 3$) and study their effects on the phase diagram topology as well as the character of the resulting phase transitions.

\begin{acknowledgments}
This work was supported by the Scientific Grant Agency of Ministry of Education of Slovak Republic (Grant No. 1/0531/19) and the Slovak Research and Development Agency (Contract No. APVV-18-0197). The authors would also like to thank the Joint Institute for Nuclear Research in Dubna, Russian Federation for the use of their Govorun Supercomputer.
\end{acknowledgments}


\begin{thebibliography}{27}
\expandafter\ifx\csname natexlab\endcsname\relax\def\natexlab#1{#1}\fi
\expandafter\ifx\csname bibnamefont\endcsname\relax
  \def\bibnamefont#1{#1}\fi
\expandafter\ifx\csname bibfnamefont\endcsname\relax
  \def\bibfnamefont#1{#1}\fi
\expandafter\ifx\csname citenamefont\endcsname\relax
  \def\citenamefont#1{#1}\fi
\expandafter\ifx\csname url\endcsname\relax
  \def\url#1{\texttt{#1}}\fi
\expandafter\ifx\csname urlprefix\endcsname\relax\def\urlprefix{URL }\fi
\providecommand{\bibinfo}[2]{#2}
\providecommand{\eprint}[2][]{\url{#2}}

\bibitem[{\citenamefont{Mermin and Wagner}(1966)}]{Mermin_Wagner_1966}
\bibinfo{author}{\bibfnamefont{N.~D.} \bibnamefont{Mermin}} \bibnamefont{and}
  \bibinfo{author}{\bibfnamefont{H.}~\bibnamefont{Wagner}},
  \bibinfo{journal}{Phys. Rev. Lett.} \textbf{\bibinfo{volume}{17}},
  \bibinfo{pages}{1133} (\bibinfo{year}{1966}),
  \urlprefix\url{https://link.aps.org/doi/10.1103/PhysRevLett.17.1133}.

\bibitem[{\citenamefont{{Kosterlitz, J. M. and Thouless, D. J.}}(1973)}]{BKT1}
\bibinfo{author}{\bibnamefont{{Kosterlitz, J. M. and Thouless, D. J.}}},
  \bibinfo{journal}{Journal of Physics C: Solid State Physics}
  \textbf{\bibinfo{volume}{6}}, \bibinfo{pages}{1181} (\bibinfo{year}{1973}),
  ISSN \bibinfo{issn}{0022-3719},
  \urlprefix\url{http://stacks.iop.org/0022-3719/6/i=7/a=010}.

\bibitem[{\citenamefont{Kosterlitz}(1974)}]{BKT2}
\bibinfo{author}{\bibfnamefont{J.~M.} \bibnamefont{Kosterlitz}},
  \bibinfo{journal}{Journal of Physics C: Solid State Physics}
  \textbf{\bibinfo{volume}{7}}, \bibinfo{pages}{1046} (\bibinfo{year}{1974}),
  \urlprefix\url{https://doi.org/10.1088%2F0022-3719%2F7%2F6%2F005}.

\bibitem[{\citenamefont{Miyashita and Shiba}(1984)}]{Miyashita_1984}
\bibinfo{author}{\bibfnamefont{S.}~\bibnamefont{Miyashita}} \bibnamefont{and}
  \bibinfo{author}{\bibfnamefont{H.}~\bibnamefont{Shiba}},
  \bibinfo{journal}{Journal of the Physical Society of Japan}
  \textbf{\bibinfo{volume}{53}}, \bibinfo{pages}{1145} (\bibinfo{year}{1984}),
  \eprint{https://doi.org/10.1143/JPSJ.53.1145},
  \urlprefix\url{https://doi.org/10.1143/JPSJ.53.1145}.

\bibitem[{\citenamefont{Lee et~al.}(1986)\citenamefont{Lee, Joannopoulos,
  Negele, and Landau}}]{Lee_1986}
\bibinfo{author}{\bibfnamefont{D.~H.} \bibnamefont{Lee}},
  \bibinfo{author}{\bibfnamefont{J.~D.} \bibnamefont{Joannopoulos}},
  \bibinfo{author}{\bibfnamefont{J.~W.} \bibnamefont{Negele}},
  \bibnamefont{and} \bibinfo{author}{\bibfnamefont{D.~P.}
  \bibnamefont{Landau}}, \bibinfo{journal}{Phys. Rev. B}
  \textbf{\bibinfo{volume}{33}}, \bibinfo{pages}{450} (\bibinfo{year}{1986}),
  \urlprefix\url{https://link.aps.org/doi/10.1103/PhysRevB.33.450}.

\bibitem[{\citenamefont{Lee and Lee}(1998)}]{Lee_1998}
\bibinfo{author}{\bibfnamefont{S.}~\bibnamefont{Lee}} \bibnamefont{and}
  \bibinfo{author}{\bibfnamefont{K.-C.} \bibnamefont{Lee}},
  \bibinfo{journal}{Phys. Rev. B} \textbf{\bibinfo{volume}{57}},
  \bibinfo{pages}{8472} (\bibinfo{year}{1998}),
  \urlprefix\url{https://link.aps.org/doi/10.1103/PhysRevB.57.8472}.

\bibitem[{\citenamefont{Korshunov}(2002)}]{Korshunov_2002}
\bibinfo{author}{\bibfnamefont{S.~E.} \bibnamefont{Korshunov}},
  \bibinfo{journal}{Phys. Rev. Lett.} \textbf{\bibinfo{volume}{88}},
  \bibinfo{pages}{167007} (\bibinfo{year}{2002}),
  \urlprefix\url{https://link.aps.org/doi/10.1103/PhysRevLett.88.167007}.

\bibitem[{\citenamefont{Hasenbusch et~al.}(2005)\citenamefont{Hasenbusch,
  Pelissetto, and Vicari}}]{Hasenbusch_2005}
\bibinfo{author}{\bibfnamefont{M.}~\bibnamefont{Hasenbusch}},
  \bibinfo{author}{\bibfnamefont{A.}~\bibnamefont{Pelissetto}},
  \bibnamefont{and} \bibinfo{author}{\bibfnamefont{E.}~\bibnamefont{Vicari}},
  \bibinfo{journal}{Phys. Rev. B} \textbf{\bibinfo{volume}{72}},
  \bibinfo{pages}{184502} (\bibinfo{year}{2005}),
  \urlprefix\url{https://link.aps.org/doi/10.1103/PhysRevB.72.184502}.

\bibitem[{\citenamefont{Obuchi and Kawamura}(2012)}]{Obuchi_2012}
\bibinfo{author}{\bibfnamefont{T.}~\bibnamefont{Obuchi}} \bibnamefont{and}
  \bibinfo{author}{\bibfnamefont{H.}~\bibnamefont{Kawamura}},
  \bibinfo{journal}{Journal of the Physical Society of Japan}
  \textbf{\bibinfo{volume}{81}}, \bibinfo{pages}{054003}
  (\bibinfo{year}{2012}), \eprint{https://doi.org/10.1143/JPSJ.81.054003},
  \urlprefix\url{https://doi.org/10.1143/JPSJ.81.054003}.

\bibitem[{\citenamefont{Lee and Grinstein}(1985)}]{Lee_Grinstein_1985}
\bibinfo{author}{\bibfnamefont{D.~H.} \bibnamefont{Lee}} \bibnamefont{and}
  \bibinfo{author}{\bibfnamefont{G.}~\bibnamefont{Grinstein}},
  \bibinfo{journal}{Phys. Rev. Lett.} \textbf{\bibinfo{volume}{55}},
  \bibinfo{pages}{541} (\bibinfo{year}{1985}),
  \urlprefix\url{https://link.aps.org/doi/10.1103/PhysRevLett.55.541}.

\bibitem[{\citenamefont{Korshunov}(1985)}]{Korshunov_1985}
\bibinfo{author}{\bibfnamefont{E.}~\bibnamefont{Korshunov}, \bibfnamefont{S}},
  \bibinfo{journal}{Journal of Experimental and Theoretical Physics}
  \textbf{\bibinfo{volume}{41}}, \bibinfo{pages}{216} (\bibinfo{year}{1985}).

\bibitem[{\citenamefont{{Sluckin, T.J.} and {Ziman,
  Timothy}}(1988)}]{Sluckin_1988}
\bibinfo{author}{\bibnamefont{{Sluckin, T.J.}}} \bibnamefont{and}
  \bibinfo{author}{\bibnamefont{{Ziman, Timothy}}}, \bibinfo{journal}{J. Phys.
  France} \textbf{\bibinfo{volume}{49}}, \bibinfo{pages}{567}
  (\bibinfo{year}{1988}),
  \urlprefix\url{https://doi.org/10.1051/jphys:01988004904056700}.

\bibitem[{\citenamefont{Carpenter and Chalker}(1989)}]{Carpenter_1989}
\bibinfo{author}{\bibfnamefont{D.~B.} \bibnamefont{Carpenter}}
  \bibnamefont{and} \bibinfo{author}{\bibfnamefont{J.~T.}
  \bibnamefont{Chalker}}, \bibinfo{journal}{Journal of Physics: Condensed
  Matter} \textbf{\bibinfo{volume}{1}}, \bibinfo{pages}{4907}
  (\bibinfo{year}{1989}),
  \urlprefix\url{https://doi.org/10.1088%2F0953-8984%2F1%2F30%2F004}.

\bibitem[{\citenamefont{Hlubina}(2008)}]{hlub08}
\bibinfo{author}{\bibfnamefont{R.}~\bibnamefont{Hlubina}},
  \bibinfo{journal}{Phys. Rev. B} \textbf{\bibinfo{volume}{77}},
  \bibinfo{pages}{094503} (\bibinfo{year}{2008}),
  \urlprefix\url{https://link.aps.org/doi/10.1103/PhysRevB.77.094503}.

\bibitem[{\citenamefont{Qi et~al.}(2013)\citenamefont{Qi, Qin, Jia, and
  Liu}}]{QI_2013}
\bibinfo{author}{\bibfnamefont{K.}~\bibnamefont{Qi}},
  \bibinfo{author}{\bibfnamefont{M.}~\bibnamefont{Qin}},
  \bibinfo{author}{\bibfnamefont{X.}~\bibnamefont{Jia}}, \bibnamefont{and}
  \bibinfo{author}{\bibfnamefont{J.-M.} \bibnamefont{Liu}},
  \bibinfo{journal}{Journal of Magnetism and Magnetic Materials}
  \textbf{\bibinfo{volume}{340}}, \bibinfo{pages}{127 } (\bibinfo{year}{2013}),
  ISSN \bibinfo{issn}{0304-8853},
  \urlprefix\url{http://www.sciencedirect.com/science/article/pii/S0304885313002163}.

\bibitem[{\citenamefont{H\"ubscher and Wessel}(2013)}]{Hubscher_2013}
\bibinfo{author}{\bibfnamefont{D.~M.} \bibnamefont{H\"ubscher}}
  \bibnamefont{and} \bibinfo{author}{\bibfnamefont{S.}~\bibnamefont{Wessel}},
  \bibinfo{journal}{Phys. Rev. E} \textbf{\bibinfo{volume}{87}},
  \bibinfo{pages}{062112} (\bibinfo{year}{2013}),
  \urlprefix\url{https://link.aps.org/doi/10.1103/PhysRevE.87.062112}.

\bibitem[{\citenamefont{Park et~al.}(2008)\citenamefont{Park, Onoda, Nagaosa,
  and Han}}]{Park-nematic}
\bibinfo{author}{\bibfnamefont{J.-H.} \bibnamefont{Park}},
  \bibinfo{author}{\bibfnamefont{S.}~\bibnamefont{Onoda}},
  \bibinfo{author}{\bibfnamefont{N.}~\bibnamefont{Nagaosa}}, \bibnamefont{and}
  \bibinfo{author}{\bibfnamefont{J.~H.} \bibnamefont{Han}},
  \bibinfo{journal}{Phys. Rev. Lett.} \textbf{\bibinfo{volume}{101}},
  \bibinfo{pages}{167202} (\bibinfo{year}{2008}),
  \urlprefix\url{https://link.aps.org/doi/10.1103/PhysRevLett.101.167202}.

\bibitem[{\citenamefont{Qin et~al.}(2009)\citenamefont{Qin, Chen, and
  Liu}}]{qin09}
\bibinfo{author}{\bibfnamefont{M.~H.} \bibnamefont{Qin}},
  \bibinfo{author}{\bibfnamefont{X.}~\bibnamefont{Chen}}, \bibnamefont{and}
  \bibinfo{author}{\bibfnamefont{J.~M.} \bibnamefont{Liu}},
  \bibinfo{journal}{Phys. Rev. B} \textbf{\bibinfo{volume}{80}},
  \bibinfo{pages}{224415} (\bibinfo{year}{2009}),
  \urlprefix\url{https://link.aps.org/doi/10.1103/PhysRevB.80.224415}.

\bibitem[{\citenamefont{Hayden et~al.}(2010)\citenamefont{Hayden, Kaplan, and
  Mahanti}}]{hayd10}
\bibinfo{author}{\bibfnamefont{L.~X.} \bibnamefont{Hayden}},
  \bibinfo{author}{\bibfnamefont{T.~A.} \bibnamefont{Kaplan}},
  \bibnamefont{and} \bibinfo{author}{\bibfnamefont{S.~D.}
  \bibnamefont{Mahanti}}, \bibinfo{journal}{Phys. Rev. Lett.}
  \textbf{\bibinfo{volume}{105}}, \bibinfo{pages}{047203}
  (\bibinfo{year}{2010}),
  \urlprefix\url{https://link.aps.org/doi/10.1103/PhysRevLett.105.047203}.

\bibitem[{\citenamefont{Dian and Hlubina}(2011)}]{dian11}
\bibinfo{author}{\bibfnamefont{M.}~\bibnamefont{Dian}} \bibnamefont{and}
  \bibinfo{author}{\bibfnamefont{R.}~\bibnamefont{Hlubina}},
  \bibinfo{journal}{Phys. Rev. B} \textbf{\bibinfo{volume}{84}},
  \bibinfo{pages}{224420} (\bibinfo{year}{2011}),
  \urlprefix\url{https://link.aps.org/doi/10.1103/PhysRevB.84.224420}.

\bibitem[{\citenamefont{\ifmmode \check{Z}\else
  \v{Z}\fi{}ukovi\ifmmode~\check{c}\else \v{c}\fi{}}(2019)}]{zuko19}
\bibinfo{author}{\bibfnamefont{M.}~\bibnamefont{\ifmmode \check{Z}\else
  \v{Z}\fi{}ukovi\ifmmode~\check{c}\else \v{c}\fi{}}}, \bibinfo{journal}{Phys.
  Rev. E} \textbf{\bibinfo{volume}{99}}, \bibinfo{pages}{062112}
  (\bibinfo{year}{2019}),
  \urlprefix\url{https://link.aps.org/doi/10.1103/PhysRevE.99.062112}.

\bibitem[{\citenamefont{Grason}(2008)}]{gras08}
\bibinfo{author}{\bibfnamefont{G.~M.} \bibnamefont{Grason}},
  \bibinfo{journal}{{EPL} (Europhysics Letters)} \textbf{\bibinfo{volume}{83}},
  \bibinfo{pages}{58003} (\bibinfo{year}{2008}),
  \urlprefix\url{https://doi.org/10.1209%2F0295-5075%2F83%2F58003}.

\bibitem[{\citenamefont{\ifmmode \check{Z}\else
  \v{Z}\fi{}ukovi\ifmmode~\check{c}\else
  \v{c}\fi{}}(2016)}]{zukovic_frustrated_2016}
\bibinfo{author}{\bibfnamefont{M.}~\bibnamefont{\ifmmode \check{Z}\else
  \v{Z}\fi{}ukovi\ifmmode~\check{c}\else \v{c}\fi{}}}, \bibinfo{journal}{Phys.
  Rev. B} \textbf{\bibinfo{volume}{94}}, \bibinfo{pages}{014438}
  (\bibinfo{year}{2016}),
  \urlprefix\url{https://link.aps.org/doi/10.1103/PhysRevB.94.014438}.

\bibitem[{\citenamefont{Poderoso et~al.}(2011)\citenamefont{Poderoso, Arenzon,
  and Levin}}]{Poderoso-2011}
\bibinfo{author}{\bibfnamefont{F.~C.} \bibnamefont{Poderoso}},
  \bibinfo{author}{\bibfnamefont{J.~J.} \bibnamefont{Arenzon}},
  \bibnamefont{and} \bibinfo{author}{\bibfnamefont{Y.}~\bibnamefont{Levin}},
  \bibinfo{journal}{Phys. Rev. Lett.} \textbf{\bibinfo{volume}{106}},
  \bibinfo{pages}{067202} (\bibinfo{year}{2011}),
  \urlprefix\url{https://link.aps.org/doi/10.1103/PhysRevLett.106.067202}.

\bibitem[{\citenamefont{Canova et~al.}(2014)\citenamefont{Canova, Levin, and
  Arenzon}}]{Canova-2014}
\bibinfo{author}{\bibfnamefont{G.~A.} \bibnamefont{Canova}},
  \bibinfo{author}{\bibfnamefont{Y.}~\bibnamefont{Levin}}, \bibnamefont{and}
  \bibinfo{author}{\bibfnamefont{J.~J.} \bibnamefont{Arenzon}},
  \bibinfo{journal}{Phys. Rev. E} \textbf{\bibinfo{volume}{89}},
  \bibinfo{pages}{012126} (\bibinfo{year}{2014}),
  \urlprefix\url{https://link.aps.org/doi/10.1103/PhysRevE.89.012126}.

\bibitem[{\citenamefont{Canova et~al.}(2016)\citenamefont{Canova, Levin, and
  Arenzon}}]{Canova-2016}
\bibinfo{author}{\bibfnamefont{G.~A.} \bibnamefont{Canova}},
  \bibinfo{author}{\bibfnamefont{Y.}~\bibnamefont{Levin}}, \bibnamefont{and}
  \bibinfo{author}{\bibfnamefont{J.~J.} \bibnamefont{Arenzon}},
  \bibinfo{journal}{Phys. Rev. E} \textbf{\bibinfo{volume}{94}},
  \bibinfo{pages}{032140} (\bibinfo{year}{2016}),
  \urlprefix\url{https://link.aps.org/doi/10.1103/PhysRevE.94.032140}.

\bibitem[{\citenamefont{Suzuki}(1974)}]{suzuki_WU}
\bibinfo{author}{\bibfnamefont{M.}~\bibnamefont{Suzuki}},
  \bibinfo{journal}{Progress of Theoretical Physics}
  \textbf{\bibinfo{volume}{51}}, \bibinfo{pages}{1992} (\bibinfo{year}{1974}),
  ISSN \bibinfo{issn}{0033-068X},
  \eprint{https://academic.oup.com/ptp/article-pdf/51/6/1992/5276871/51-6-1992.pdf},
  \urlprefix\url{https://doi.org/10.1143/PTP.51.1992}.

\end{thebibliography}

\end{document}